\pdfoutput=1
\documentclass[usenatbib]{mnras}

\usepackage{mathptmx}
\usepackage[british]{babel}
\usepackage{ifthen}
\usepackage{etoolbox}
\usepackage{graphicx} 
\usepackage{microtype}
\usepackage{xspace}
\usepackage{amssymb,amsmath}
\usepackage{paralist}
\usepackage{placeins}
\usepackage{float}
\usepackage{enumitem}
\usepackage{multirow}
\usepackage{bm}
\usepackage{breqn}
\usepackage[T1]{fontenc}

\newcommand{\ie}{{\it i.e.}\xspace}
\newcommand{\eg}{{\it e.g.}\xspace}
\newcommand{\cf}{{\it cf.}\xspace}
\newcommand{\etal}{{\it et al.}\xspace}
\newcommand{\kpc}{\ensuremath{\,{\rm kpc}}\xspace}
\newcommand{\pc}{\ensuremath{\,{\rm pc}}\xspace}

\newcommand{\Gyr}{\ensuremath{\,{\rm Gyr}}\xspace}
\newcommand{\kms}{\ensuremath{\,{\rm km}\,{\rm s}^{-1}}\xspace}
\newcommand{\mags}{\ensuremath{\,{\rm mag}}\xspace}
\newcommand{\days}{\ensuremath{\,{\rm days}}\xspace}

\newcommand{\I}{\ensuremath{{I}}\xspace}

\newcommand{\Iband}{\I-band\xspace}

\newcommand{\msun}{\ensuremath{{M_\odot}}\xspace}

\newcommand{\rcut}{\ensuremath{R_{\rm cut}}\xspace}

\newcommand{\logten}{{\log~}}

\newcommand{\twosigmax}{72\%\xspace}
\newcommand{\fv}{{\ensuremath{f_v}\xspace}}
\newcommand{\onesigbounds}{{\ensuremath{f_v=(0.88\pm0.07)}\xspace}}
\newcommand{\onesigboundseros}{{\ensuremath{f_v=(0.9\pm0.1)}\xspace}}
\newcommand{\twosigmaxfrac}{{\ensuremath{f_v > 0.72}\xspace}}

\newcommand{\ds}{\ensuremath{D_{\rm s}}\xspace}
\newcommand{\dl}{\ensuremath{D_{\rm l}}\xspace}
\newcommand{\di}{\ensuremath{D_i}\xspace}
\newcommand{\rhol}{\ensuremath{\rho _{\rm l}}\xspace}
\newcommand{\rhos}{\ensuremath{\rho _{\rm s}}\xspace}
\newcommand{\te}{\ensuremath{t_{\rm E}}\xspace}

\title[MOA-II Galactic Microlensing Constraints]{MOA-II Galactic Microlensing Constraints: The Inner Milky Way has a Low Dark Matter Fraction and a Near Maximal Disk}
 
\author[Wegg \etal]
{Christopher Wegg$^{1}$\thanks{E-mail: wegg@mpe.mpg.de}, Ortwin Gerhard$^{1}$
and Matthieu Portail$^{1}$\\
$^1$Max-Planck-Institut f\"ur Extraterrestrische Physik, Giessenbachstrasse,
85748 Garching, Germany.}
\pagerange{\pageref{firstpage}--\pageref{lastpage}} \pubyear{2015}
 
\begin{document}
\label{firstpage}
\maketitle

\begin{abstract}

Microlensing provides a unique tool to break the stellar to dark matter degeneracy in the inner Milky Way. We combine N-body dynamical models fitted to the Milky Way's Boxy/Peanut bulge with exponential disk models outside this, and compute the microlensing properties. Considering the range of models consistent with the revised MOA-II data, we find low dark matter fractions in the inner Galaxy: at the peak of their stellar rotation curve a fraction \onesigbounds~of the circular velocity is baryonic (at $1\sigma$, \twosigmaxfrac~at $2\sigma$). These results are in agreement with constraints from the EROS-II microlensing survey of brighter resolved stars, where we find \onesigboundseros~at $1\sigma$.  Our fiducial model of a disk with scale length 2.6\kpc, and a bulge with a low dark matter fraction of 12\%, agrees with both the revised MOA-II and EROS-II microlensing data. The required baryonic fractions, and the resultant low contribution from dark matter, are consistent with the NFW profiles produced by dissipationless cosmological simulations in Milky Way mass galaxies. They are also consistent with recent prescriptions for the mild adiabatic contraction of Milky Way mass haloes without the need for strong feedback, but there is some tension with recent measurements of the local dark matter density. Microlensing optical depths from the larger OGLE-III sample could improve these constraints further when available.

\end{abstract}
\begin{keywords}
Galaxy: bulge -- Galaxy: center -- Galaxy: structure -- gravitational lensing:
micro.
\end{keywords}

\section{Introduction}

The concept of microlensing, whereby a background star is gravitationally lensed by a foreground object, was conceived by \citet{Einstein:36} however it was not until \citet{Paczynski:86} that modern microlensing experiments were born. 

One of the primary observational microlensing measurements is the optical depth. The optical depth is defined as the fraction of stars whose projection lies within the Einstein radius of a lens, and it is these stars who see a significant ($>3/\sqrt{5}$) brightening due to microlensing. For a star at distance \ds the optical depth $\tau$ is given by \citep[\eg][]{Kiraga:94}
\begin{equation}
	\tau(\ds) = \frac{4\pi G }{c^2} \int_0^\ds \rhol(\dl) \left(\frac{1}{\dl} - \frac{1}{\ds} \right) \dl^2 \, d\dl  \label{eq:tau} \end{equation}
where $\rhol(\dl)$ is the density of lenses. That the optical depth depends only on the density of lenses, and not on their distribution of masses or velocities, makes it theoretically very elegant.

An original motivation for microlensing studies was to explore the possibility that the Galactic dark halo could be composed of Massive Astrophysical Compact Halo Objects (MACHOs) \citep{Paczynski:86}.   The first results from the MACHO survey measuring the optical depth towards the LMC suggested, based on $\sim 10$ events, that 20\% of the Galactic halo could be composed of MACHOs with average mass $\sim 0.4\msun$ \citep{Alcock:00,Bennett:05}. In contrast the EROS survey found that less than 8\% of the halo could be composed of MACHOs with mass $\sim 0.4\msun$ \citep{Tisserand:07}. The OGLE-III survey was much larger in both duration and coverage and found stringent limits of the contribution of MACHOs: a fraction less than $7\%$ for sub-solar lenses \citep{Wyrzykowski:11}. Hence, the optical depth observed towards the LMC has turned out to be too small for the halo to contain more than a small fraction of dark objects able to microlense, most likely because dark matter is composed of particles with low mass, so that they have an Einstein radius far too small to produce a significant brightening  \citep[\ie $\lesssim 10^{-7}\msun$, ][]{Paczynski:86}.  

This instead makes microlensing of Milky Way bulge stars a unique tool for unravelling the structure of galaxies because only stellar matter is able to produce microlensing events. Typically the mass profiles of Galaxies can be constrained dynamically, but this may be distributed between baryonic and dark matter \citep[\eg][]{Courteau:14}. In contrast we are able to use bulge microlensing to break the stellar to dark matter degeneracy in the inner Milky way ($R \lesssim 5\kpc$) in a manner that is generally not possible in external galaxies without additional assumptions such as on the stellar mass-to-light ratio \citep{Iocco:11}.

Microlensing towards the bulge was first considered by \citet{Paczynski:91} and \citet{Griest:91bulge}. Initial predictions were based on bulge models fitted to COBE data \citep{Han:95,Bissantz:97,Evans:02,Bissantz:02}. However the measured optical depths were significantly higher than these models predicted. OGLE measured $\tau=(3.3\pm1.2)\times 10^{-6}$ \citep{Udalski:94} and MACHO $\tau =(3.9\substack{+1.8 \\ -1.2})\times 10^{-6}$ \citep{Alcock:97}. These early optical depths were higher even than the theoretical bounds: no model could be constructed that reproduced $\tau$ without overshooting the rotation curve \citep{Binney:00}. Later estimates with improved treatment of blending through difference image analysis (DIA) \citep{Alcock:00bulge} and focusing on brighter resolved giants \citep{Popowski:05,Hamadache:06} produced reduced optical depths, so that at least some possible Galactic models were consistent with the data \citep[\eg][]{Wood:05}, and actually brought the optical depth into line with the earlier predictions.

Pioneering studies from OGLE \citep{Udalski:94} and MACHO \citep{Alcock:97} were based on a handful of microlensing events. Since then several thousand microlensing events have been detected by  recent and ongoing microlensing surveys such as EROS-2 \citep{Afonso:03,Hamadache:06}, MOA-2 \citep{Sumi:11,Sumi:13,Sumi:16}, MACHO \citep{Popowski:05}, WeCAPP \citep{Lee:15}, OGLE-III \citep{Wyrzykowski:15}, and OGLE-IV \citep{Udalski:15}. These larger samples, the convergence between measurements  of recent optical depths for bright resolved stars with fainter unresolved stars \citep{Sumi:16}, and the new made-to-measure models of the bulge \citep{Portail:15}, prompts us to revisit of the constraints provided by bulge microlensing data on Galactic models. We use recent microlensing data to constrain the amount of stellar matter towards the Galactic bulge, and consequently the  fractional contributions of stellar and dark matter in the inner Galaxy. The data is primarily taken from the MOA-II survey \citep{Sumi:13} which was recently updated by \citet{Sumi:16} with improved estimates of the effective number of monitored stars. We use the data from \citet{Sumi:16} throughout, referring to it as the revised MOA-II data. We additionally consider and cross check our results with data from the EROS-II survey \citep{Hamadache:06}.

\begin{table*}
\caption{Summary description of the model Galaxy used to predict microlensing quantities.}
\label{tab:model}
\begin{tabular}{llll}
\hline
& Property & Fiducial & Variations \\
\hline
Bulge Model & N-body model from \citet{Portail:15} & M90 \ & M80, M85 \\
\hline
\multirow{4}{*}{Disk Model} & Disk scale length, $R_d$ & 2.6\kpc & Range: $1.9-3.4\kpc$\\
& Solar neighbourhood disk scale height, $H_\odot$ & 0.3\kpc & --- \\
& Inner disk scale height , $H_{4.5}$ & 0.18\kpc & 0.3\kpc \\
& Local stellar surface density, $\Sigma_\odot$ & $38\, \msun\,\kpc^{-2}$ & --- \\
\hline
\multirow{4}{*}{Stellar Population} & Source stellar population & 10Gyr, Baade's Window MDF from \citet{Zoccali:08} & --- \\
& Isochrones & $\alpha$-enhanced BASTI \citep{Pietrinferni:04} & --- \\
& IMF & \citet{Kroupa:01} & \citet{salpeter:55} \\
& Remnant masses & Prescription from \citet{Maraston:98} & --- \\
\hline
\end{tabular}
\end{table*}

This paper proceeds as follows:
In \autoref{sec:models} we describe our models of the distribution of stellar matter in the Milky Way.
In \autoref{sec:microtheory} we describe how we calculate microlensing properties such as the optical depth and timescale distribution which can then be subsequently compared to the data in sections \ref{sec:data} and \ref{sec:eros}. We compute the resultant rotation curves consistent with data in \autoref{sec:rotcurve}, and briefly investigate the timescale distribution of events in \autoref{sec:tedistdiscuss}. We discuss the consequences of the results in \autoref{sec:discuss}, in particular the limits these place on the halo contribution, and place our Milky Way constraints in the context of external galaxies. We conclude in \autoref{sec:conc}.



\section{Model}
\label{sec:models}

Using the microlensing data we constrain the distribution of stellar matter in a model Milky Way galaxy. Our model consists of N-body bulges together with parametric exponential stellar disks. The properties of these models are summarised in \autoref{tab:model}.

\subsection{Bulge}
\label{sec:bulgemodels}

In the central region we use the N-body models constructed by \citet{Portail:15}. These models were constructed by the made-to-measure method to match the three dimensional density of red clump stars measured by \citet{Wegg:13} together with the radial velocity measurements provided by the BRAVA survey \citep{Kunder:12}. This combination of data highly constrains the total mass in a central box of size $(\pm 2.2 \times \pm 1.4 \times \pm 1.2 \kpc)$ to be $(1.84\pm 0.07 )\times 10^{10} \msun$. The proportions of stellar and dark matter to this total is however not well dynamically constrained. The models have stellar masses in the range $(1.35 - 1.6)\times 10^{10}\msun$. We use the same naming convention as \citet{Portail:15}, the models are labeled M80-M90 corresponding to the initial levels of disk contribution of 80-90\%  at $R=2.2R_d$. 

As our fiducial model we use model M90, and consider the other variants, M80 and M85, when noted. As described in \citet{Portail:15}, these models are expected to cover the range of possible stellar masses for the bulge on the basis of the stellar population and mass-to-light ratio for reasonable IMFs. We use model M90 as our fiducial model because it is consistent with the mass-to-light and mass-to-red clump ratios of a 10Gyr old Kroupa IMF. As we show in \autoref{sec:tedistdiscuss}, a Kroupa IMF also results in a timescale distribution of microlensing events close to that observed, making this model the most self-consistent.

\subsection{Disk}
\label{sec:diskmodels}

The N-body models of \citet{Portail:15} were fitted to the bulge where they are expected to be accurate dynamical models. Outside the bulge however they were not constrained, and therefore do not there accurately represent the Milky Way. In particular they have a disk scale length of $1.1-1.2\kpc$, significantly shorter than measured in the Milky Way \eg  typical values for inner MW disk based on NIR and surface mass density are 2.5\kpc: \citet{Binney:97}; 2.4\kpc: \citet{Bissantz:02}; 2.6\kpc: \citet{Juric:08}; 2.15\kpc: \citet{Bovy:13}; see \citet{Ortwin:araa} for a fuller discussion. This is because the self-consistent pure N-body models from which they were derived have large bar to disk scale length ratios, and therefore when the bar is scaled to the Milky Way short disk scale lengths and very low stellar densities in the solar neighbourhood result. We therefore truncate the N-body model at a radius \rcut, using the N-body models inside, and a simple analytic disk model outside.

We consider the stellar disk as a double exponential
\begin{equation}
\rho = \frac{\Sigma_\odot}{2 H(R)} \exp \left( \frac{R_0 - R}{R_d} \right) \exp \left( - \frac{|z|}{H(R)} \right)
\end{equation}
where $R$ is the galactocentric radius, $R_0$ is the solar galactocentric radius, $R_d$ is the disk scale length, $z$ the distance from the Galactic plane, and $H$ the disk scale height. 

$\Sigma_\odot$ is the stellar surface density in the solar neighbourhood and $H_\odot$ the scale height in the solar neighbourhood. The local disk is relatively well constrained and here we use fiducial local disk parameters of $\Sigma_0=38\, \msun\,\kpc^{-2}$ \citep{Bovy:13} and $H_\odot=0.3\kpc$ \citep{Juric:08}. 

The disk towards the inner Galaxy is considerably less well constrained, both in terms of its increase in surface density (parameterised here though the disk scale length $R_d$) and the disk scale height.
There is evidence that the Milky Way disk scale height may decrease inside the solar circle (\citealt{Kent:91,LopezCorredoira:02}, the 180\pc thin bar in \citealt{Wegg:15}). We therefore allow a flaring of the disk between $4.5\kpc$ and the solar neighbourhood parameterised through
\begin{equation}
	H(R)=
\begin{cases}
H_\odot + (R - R_0)\frac{H_\odot - H_{4.5}}{R_0 - 4.5\kpc} & \mbox{if } R > 4.5\kpc \\
H_{4.5} & \mbox{if } R \le 4.5\kpc
\end{cases} ~ .
\end{equation}
We consider two choices: either $H_{4.5}=180\pc$ \citep[the long bar scale height found in][]{Wegg:15}, or that the scale height is constant and $H_{4.5}=300\pc$.

We choose this simple disk model since the microlensing results are most sensitive to the stellar disk inwards towards the bulge, where the disk density and profile is highly uncertain, and therefore more complex models are not justified.

As our fiducial value of $R_d$ we use 2.6\kpc \citep{Juric:08}. When considering variations on this we use a range of 1.9\kpc to 3.4\kpc considering this to cover the range deemed reasonable by recent data  \citep{Ortwin:araa}. 

We switch between the disk and bulge models at a radius \rcut which is chosen so that the azimuthally averaged surface density is continuous. This choice is made so that the rotation curve is smooth. A typical value of \rcut is $1.6\kpc$ from the fiducial model.

When velocities of the disk are required we use a Schwarzschild velocity distribution with mean velocity taken from the rotation curve of \citet{Sofue:09}. For the local velocity dispersions we use values from \citet{Dehnen:1998} and extrapolate these inwards using an exponential increase $\exp (-R/R_\sigma)$ with $R_\sigma=2\times4.37\kpc$ \citep{Lewis:89}. Our results are not sensitive to these choices. This is because they are based on the optical depth which is not highly sensitive to the velocities and resultant timescale distribution as shown by the IMF independence in \autoref{sec:minoraxis}.

Although we use an analytic model for the disk, in practice we produce N-body realisations when computing microlensing quantities for consistency with the N-body bulge model.

\subsection{Stellar Population}
\label{sec:pop}

Because the majority of stars microlensed are in the bulge, when the luminosity function is required we use the $\alpha$-enhanced BASTI isochrones from \citet{Pietrinferni:04} for 10Gyr, the MDF measured by \citet{Zoccali:08} towards Baade's window, and a \citet{Kroupa:01} IMF. We show in \autoref{sec:minoraxis} the effect of altering the IMF to the \citet{salpeter:55} IMF and consider the effects of the IMF on the timescale distribution in \autoref{sec:tedistdiscuss}. For the lens mass distribution a remnant mass prescription is required. For this we use the same prescription for white dwarfs, black holes and neutron stars as \citet{Maraston:98}: White dwarfs are formed from initial masses $M_{\rm i}<8.5\msun$ with remnant mass $M_{\rm f}=0.077M_{\rm i}+0.48\msun$, neutron stars of mass $M_{\rm f}=1.4\msun$ result from $8.5\msun \le M_{\rm i} < 40\msun$, and black holes of mass $M_{\rm f}=0.5 M_{\rm i}$ result from $M_{\rm i} \ge 40\msun$. Throughout we assume that white dwarfs, black holes and neutron stars receive no kick at birth and are distributed as their progenitors.

\section{Microlensing Parameters}
\label{sec:microtheory}
\subsection{Optical Depth}
\label{sec:tautheory}

In our N-body models, for a source particle $i$ at distance $\di$ the optical depth $\tau_i$ can be calculated as a Monte-Carlo integral  over the foreground lens particles in a small field around the star of interest (\cf \autoref{eq:tau})
\begin{equation}
	\tau_i = \frac{4\pi G}{c^2 \omega} \sum_j M_j \left(\frac{1}{D_j} - \frac{1}{D_i} \right) \label{eq:taumc}
\end{equation}
where $M_j$ is the mass of particle $j$ and the sum runs over all particles in a pencil beam of solid angle $\omega$ with distance $D_j<D_i$. We use this notation throughout.

The values of optical depth measured by observational surveys however are an average over observable stars. Typically two values are given: An optical depth for all sources, and one confined to the brighter giants in the region of the colour-magnitude diagram which contains red clump giants (RCGs). These two values can differ both because surveys may have problems with blending of unresolved fainter stars which is less important for the brighter RCGs, and also because their luminosity function and therefore weighting by distance is different. Since RCGs are bright enough to be observed throughout the bulge then it is typically assumed that there is no weighting by distance. For the all-source microlensing a slightly different weighting is typically utilised. Intrinsically fainter stars are more numerous than their brighter counterparts and therefore in a magnitude limited sample of bulge giants the more distant stars are down weighted by $\approx r^{-1.4}$ \citep{Portail:15}. \cite{Kiraga:94} for example parameterise this weighting by as a power law luminosity function $L^\beta$ so that the number of detectable stars in a field brighter than some faint magnitude limit is
\begin{equation}
	dn \propto \rhos(\ds) \ds^2 \ds^{2\beta} \, d\ds ~.
	\label{eq:betaparam}
\end{equation}
Stars visible throughout the bulge therefore have $\beta=0$ while other bulge giants have $\beta\approx -0.7$.

From \autoref{eq:tau} the measured optical depths are therefore the weighted average over observable stars
\begin{dmath}
	\langle \tau(\beta) \rangle = \frac{\int \rhos(\ds) \ds^{2+2\beta} \tau(\ds) \, d\ds}{\int \rhos(\ds) \ds^{2+2\beta} \, d\ds} ~.\label{eq:tauavg} 
\end{dmath}

Practically this can be calculated by taking the weighted average of \autoref{eq:taumc} \citep{Fux:97a}
\begin{dmath}
	\langle \tau(\beta) \rangle = \frac{4\pi G}{c^2 \omega} \left( \sum_i M_i D_i^{2\beta} \right)^{-1} \sum_{i,j} M_i M_j D_i^{2\beta} \left(\frac{1}{D_j} - \frac{1}{D_i} \right) \label{eq:tauavgmc} ~.
\end{dmath}

However we now show that the simple power law parametrisation \autoref{eq:betaparam} does not adequately capture the variation of $\tau$ because the luminosity function is not sufficently close to a power law. This was demonstrated by \citet{Stanek:95} and \citet{Wood:07} for source stars with magnitudes similar to bulge RCGs, and is also true near the main sequence turnoff.

An improvement over \autoref{eq:betaparam} is to use a luminosity function constructed from isochrones. In this case the number of stars with distances between $\ds$ and $\ds+d\ds$ and between extinction corrected apparent \Iband magnitude $I_0$ and $I_0+dI_0$ will be proportional to 
\begin{equation}
	dn \propto \rhos(\ds) \ds^2 \Phi_{\rm I}(I_0-5\log[\ds/10{\rm pc}]) \, d\ds \, d\I_0 ~,
\end{equation}
where $\Phi_{\rm I}$ is the \Iband luminosity function (\ie the number of stars per unit mass with absolute \Iband magnitudes between $M_I$ and $M_I+dM_I$ is $\Phi_{\rm I}(M_I) \, dM_I$). The optical depth as a function of magnitude is therefore
\begin{equation}
	\langle \tau(I_0) \rangle = \frac{\int \tau(\ds) \rhos(\ds) \Phi_{\rm I}(I_0-5\log[\ds/10{\rm pc}]) \ds^2 \, d\ds}{\int \rhos(\ds) \Phi_{\rm I}(I_0-5\log[\ds/10{\rm pc}]) \ds^2 \, d\ds} ~. 
\end{equation}
This can be written in terms of distance modulus $\mu$ as
\begin{equation}
\langle \tau(I_0) \rangle = \frac{\int \tau(\mu) \Delta(\mu) \Phi_{\rm I}(I_0-\mu) \, d\mu}{\int \Delta(\mu) \Phi_{\rm I}(I_0-\mu) \, d\mu}  \label{eq:tauvsI0}
\end{equation}
where $\Delta(\mu)=\rhos \ds^3$ written as a function of $\mu$. If the optical depth as a function of extincted source magnitude $I=I_0+A_I(D_s)$ is required, then this expression becomes
\begin{equation}
\langle \tau(I) \rangle = \frac{\int \tau(\mu_I) \Delta_I(\mu_I) \Phi_{\rm I}(I-\mu_I) \, d\mu_I}{\int \Delta_I(\mu_I) \Phi_{\rm I}(I-\mu_I) \, d\mu_I} \label{eq:tauvsI}
\end{equation}
where $\mu_I=\mu+A_I$ and
\begin{equation}
	\Delta_I(\mu_I) = \frac{\rhos \ds^3}{\left[1+ \frac{dA_I}{d\log D_s} \right]} ~. \label{eq:deltaI}
\end{equation}

We use the expressions in Equations \ref{eq:tauvsI0}--\ref{eq:deltaI} to calculate optical depths that are more accurate and representative of the data. An alternative approach is to produce Monte Carlo realisations of mock fields like the Besan\c{c}on model \citep{Robin:03} and simulate the microlensing properties of these \citep{Kerins:09,Penny:13,Awiphan:16}. In this work however we wish to consider a large range of Galactic models, and compute their microlensing properties, to decide which are consistent with the microlensing data. For this we use the fast but accurate method provided above.

For the brightest stars we expect departures from these expressions because of the finite source effect: if the magnification changes significantly over the angular diameter of the source, then the overall magnification is smaller than the point source approximation used in \autoref{eq:tau}, and therefore the optical depth is smaller. The brightest bulge giants are also the largest and so would most significant for these. A typical RCG in the bulge  however has angular radius 6$\,\mu$as, more than an order of magnitude smaller than the typical Einstein radius of bulge microlensing events. Finite source effects are there important only for the rare highest magnification events of giants, and do not significantly effect the overall optical depths for RCGs, and are negligible for MS stars.

\begin{figure}
\includegraphics[width=\linewidth]{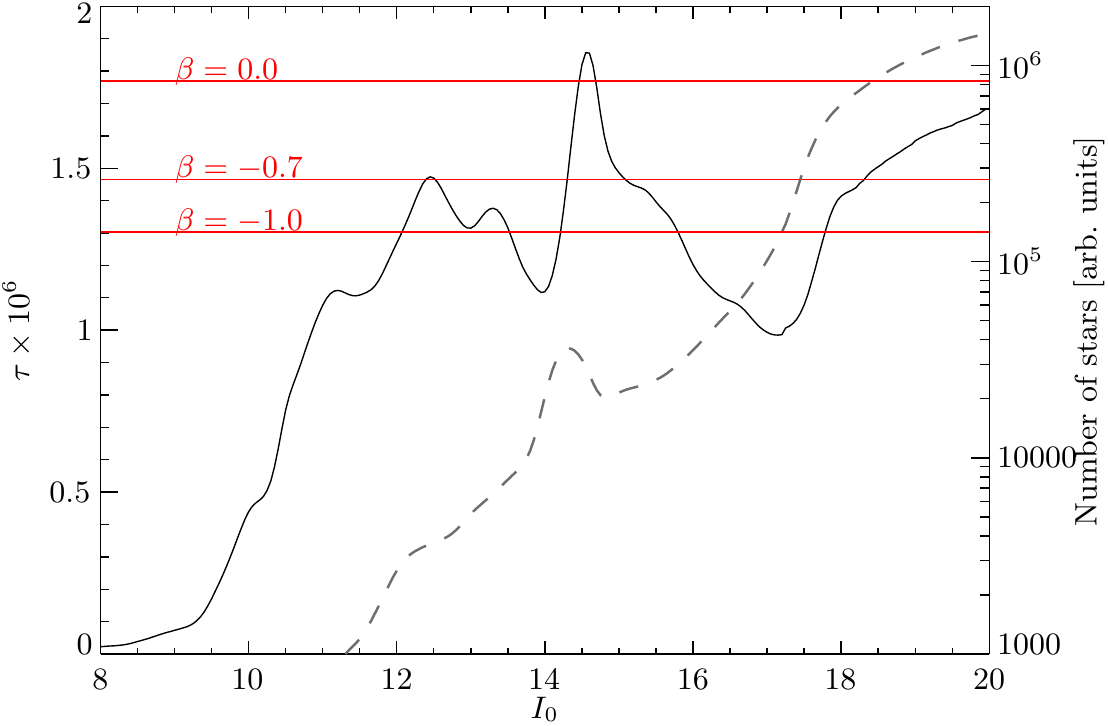}
\caption{\label{fig:tauvsImag} The microlensing optical depth as function of \Iband magnitude in the direction of Baade's window ($l=1.0^\circ$, $b=-3.9^\circ$). The optical depth of our fiducial model is shown as the solid black line calculated though \autoref{eq:tauvsI0}. The labeled red lines are the optical depth parameterised through the $\beta$ parameter as in \autoref{eq:tauavg}. The dashed grey line is the magnitude distribution predicted by the model in this field plotted against the right axis.} 
\end{figure}

Practically for the N-body models considered we compute \autoref{eq:tauvsI0} from the particles with the following procedure: up to an unimportant normalisation $\Delta(\mu)$ is just the particle masses, $M_i$, binned in equal distance modulus increments. The denominator is computed by convolving this with the luminosity function through an FFT. The numerator is computed similarly by convolution with the luminosity function after binning $M_i \tau_i$ in equal distance modulus increments. This procedure is efficient, and accurate provided that the bin size is chosen sufficiently narrow to capture the smallest scale variations in the luminosity function and distance distribution. \autoref{eq:tauvsI} can be similarly computed with each particle mass weighted by the additional denominator in \autoref{eq:deltaI}. We show in \autoref{fig:tauvsImag} a representative plot of the optical depth as a function of \Iband magnitude towards Baade's window.

Several features are clear regarding the optical as function of magnitude in \autoref{fig:tauvsImag}: For bright stars the optical depth is low. This is because, brighter than the tip of the red giant branch of the bulge, the observed stars are disk stars with low optical depth. The prominent dip, and subsequent rise in optical depth at $I_0 \approx 14$ is due to the numerous RCGs, which can also be seen in the luminosity function. Because RCGs are approximate standard candles, stars towards the bright end of the bump in the luminosity function are likely to be towards the front of the bulge, and subsequently have a low optical depth. By contrast stars towards the faint side of the bump are more likely to lie at the distant side of the bulge and therefore have a high optical depth. Fainter than this the optical depth drops, continuing to drop as disk main sequence stars which have low optical depth become numerous. The optical depth finally rises again near $I_0 \approx18$ when the main sequence turn off of the bulge is reached, and bulge dwarfs dominate the sample.

\begin{figure}
\includegraphics[width=\linewidth]{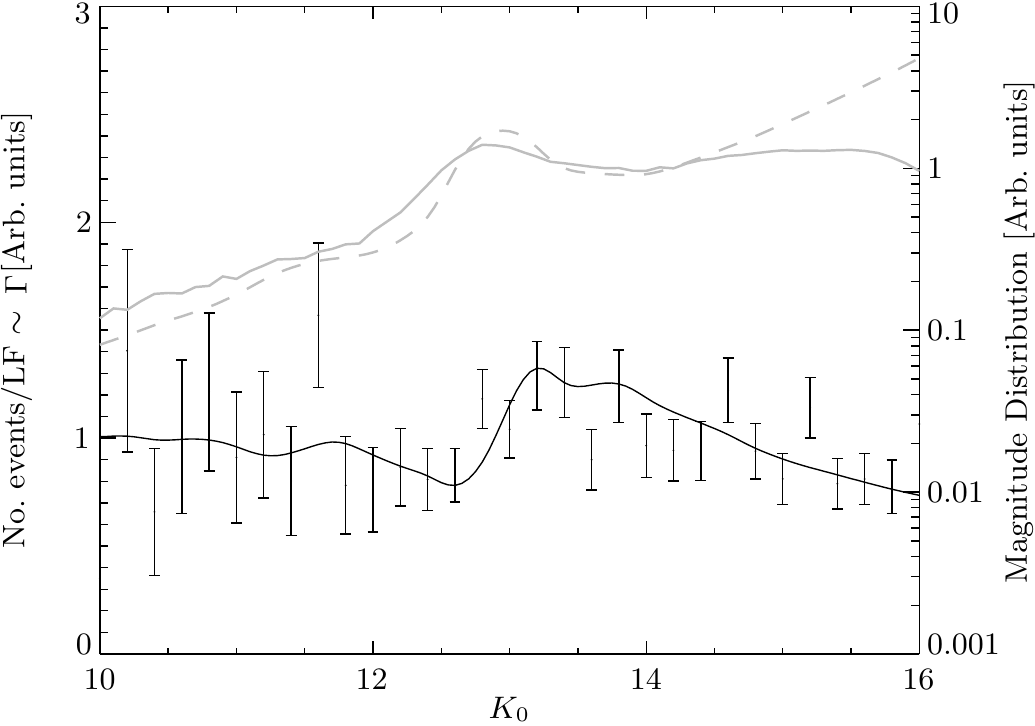}
\caption{The data with error bars are an approximate number of microlensing events per star detected by the OGLE-III survey \citep{Wyrzykowski:15} as function of extinction corrected $K_s$-band magnitude. The extinction corrected K-band magnitude distribution of microlensing events is divided by the magnitude distribution of all stars. Since the efficiency of microlensing detection changes slowly with magnitude then variations are a proxy for the change in  optical depth with magnitude. We plot as the solid black line predicted shape from our fiducial model towards Baade's window. On the right axis we plot in grey the magnitude distribution predicted by the model (dashed line) together with the magnitude distribution observed (solid line) to illustrate the position of the bump due to RCGs in the bulge and the relative position of the jump in rate. All curves were arbitrarily renormalised because the field-by-field microlensing efficiencies and number of monitored sources is not yet known for the OGLE-III survey. \label{fig:tauvsKmagOGLE}}
\end{figure}

The feature at $I_0 \approx 14$ was predicted by \citet{Stanek:95} and elaborated on by \citet{Wood:07}. We demonstrate in \autoref{fig:tauvsKmagOGLE} that it is also present in the current data. We use the OGLE-III data because it is has a higher statistical significance being a larger sample than the MOA-II sample. However the full efficiencies of the large OGLE-III sample of microlensing events has not yet been computed. Instead in this figure we plot a number which should approximately track the changes in optical depth: the number of events, divided by the magnitude distribution, as a function of extinction corrected $K_s$-band magnitude in 0.2\mags bins. The resultant number should be similar to the event rate per star, which is proportional to optical depth for a constant timescale distribution (\autoref{sec:tetheory}).

The $K$-band data is taken from VVV by cross matching to the OGLE-III events. Extinction is corrected using the $H-K_s$ colour \citep[NICE method,][]{Lada:94} and a \citet{Nishiyama:09} extinction law. The extinction corrected magnitude distribution was computed using the 100 nearest stars to each microlensing event so that it is representative. Because the efficiency changes slowly with magnitude the resultant number should change in proportion to the event rate per star, at least over narrow magnitude ranges like the feature at the magnitude of bulge red clump giants. This feature is visible in the data as the increase in event rate at $K_0 \approx 13$. 

Because of the changes in optical depth as a function of magnitude, and the inability of a single $\beta$ value to capture these changes, we urge the use of more accurate expressions such as those in Equations \ref{eq:tauvsI0}--\ref{eq:deltaI} when comparing to microlensing data.

Unfortunately a significant contribution to the optical depth arises from long duration microlensing events which are difficult to detect. As a result the error on the total microlensing optical depth can be strongly influenced by these long events. Instead what is often measured are quantities such as $\tau_{200}$, the optical depth excluding events with timescale longer than 200 days. This is also the case for the optical depth of the revised MOA-II data \citet{Sumi:16}. We therefore proceed to estimate the timescale distribution of microlensing events.

\subsection{Event Timescales and Microlensing Rates}
\label{sec:tetheory}

The event timescale is important to consider for two reasons.  First in the lowest order description of microlensing the measurable parameters for each event are the source magnitude, and the event amplification, and its duration. While the amplification is geometrical and provides no useful information on the Galaxy, the timescale depends on the relative velocity of the lens and source together with the mass of the lens and therefore has physical importance. It has for example therefore been used to provide constraints on the mass spectrum of lenses \citep[\eg][]{Han:96,CalchiNovati:08}, and low mass objects because these generate characteristically short timescale events \citep{Sumi:11}. Secondly, as mentioned above, microlensing surveys are of finite duration and therefore insensitive to long time scale events. Because these events contribute a significant amount of the total optical depth it is better to consider the optical depth excluding long duration events. The disadvantage is that both the lens mass distribution and a dynamical model is required.  

The timescale $t_E$ of a microlensing event is given by
\begin{align}
	\te &=R_E/{\rm v} = \frac{1}{\rm v} \sqrt{\frac{4 G M_{\rm l} \dl^2}{c^2} \left(\frac{1}{\dl} - \frac{1}{\ds} \right)} \\
	&=40 \, {\rm days} \left(\frac{200\kms}{\rm v}\right) \left(\frac{M}{M_\odot}\right)^{1/2} \left( \frac{\ds}{10 \kpc} \right)^{1/2}  \frac{ \left[ x(x -1)  \right]^{1/2} }{0.5} \nonumber
 \end{align}
where $R_E$ is the Einstein radius, $M_{\rm l}$ the mass of the lens, $x\equiv \dl/\ds$, and $v$ is the transverse velocity of the lens relative to the line of sight toward the source star. It is given by
\begin{equation}
	\bld{v} = \bld{v}_{\rm l} - \bld{v}_{\rm o}+(\bld{v}_{\rm o} - \bld{v}_{\rm s})\frac{\dl}{\ds}
\end{equation}
where $\bld{v}_{\rm l}$, $\bld{v}_{\rm o}$ and $\bld{v}_{\rm s}$ are the transverse velocities of the lens, observer and source respectively. 

We can calculate the event time distribution for a mono-mass lens population by considering that the optical depth is \citep{Paczynski:91}
\begin{equation}
	\tau=\frac{\pi}{2}\int \te \Gamma(\log \te) \, d\log \te~,
	\label{eq:gammadef}
\end{equation}
where $\Gamma(\log \te)$ is the event rate as a function of $\log \te$ \ie the rate of microlensing events in the range $\log \te \rightarrow \log \te+d\log\te$ is $\Gamma(\log \te)$.

Therefore, in the $\beta$ parameterisation of the luminosity function, to calculate the event time distribution from the N-body model we calculate for each lens-source pair $(j,i)$ in a field the event time, $t_{E,ij}$, and give each the weight
\begin{equation}
	d\Gamma_{ij}= \frac{2}{\pi} \frac{\tau_i}{t_{E,ij}} \frac{M_i D_i^{	2\beta}}{\sum_i M_i D_i^{2\beta}} ~.
	\label{eq:dgammaij}
\end{equation}
Binning these weights as a function of $t_E$ gives the timescale distribution.

To extend to a mass spectrum then consider a lens mass distribution  $\Phi(\logten M_l)$. The corresponding probability of an star being microlensed by mass $M_l$ at a point in time is $\propto R_E^2 \Phi(\logten M_l)  \propto M_l \Phi(\logten M_l)$. To compute the event time distribution we can first evaluate the event time distribution for a $1\msun$ population, defining this $\gamma(\logten \te)$. The multi-mass event distribution can then be deduced by considering that the event timescale is proportional to $\sqrt{M_l}$ with weighting $1/\sqrt{M_l} \times M_l \Phi(\logten M_l)$. The additional $1/\sqrt{M_l}$ weighting arises because the rate is related to the optical depth through an additional factor of $1/\te$ (\autoref{eq:gammadef}).

\begin{figure}
\centering
\includegraphics[width=0.9\linewidth]{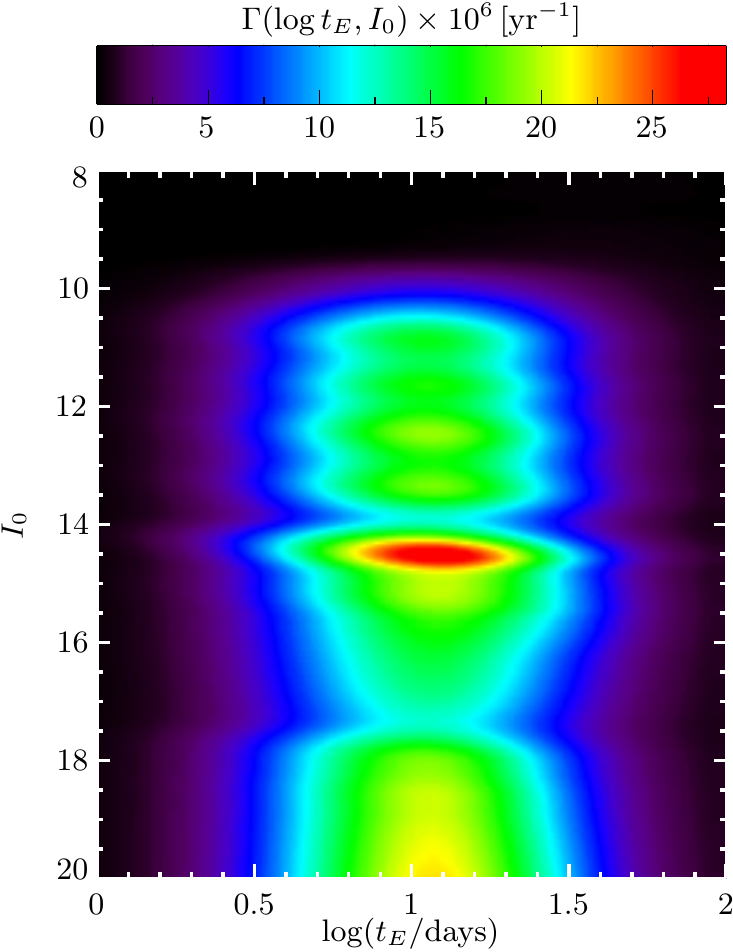}
\caption{The microlensing event rate per star as function of \Iband magnitude and event timescale $\log t_E$ in the direction of Baade's window ($l=1.0^\circ$, $b=-3.9^\circ$). The same assumptions on the stellar population as in \autoref{fig:tauvsImag} was used.  \label{fig:bw2deventrate}}
\end{figure}

\begin{figure}
\includegraphics[width=\linewidth]{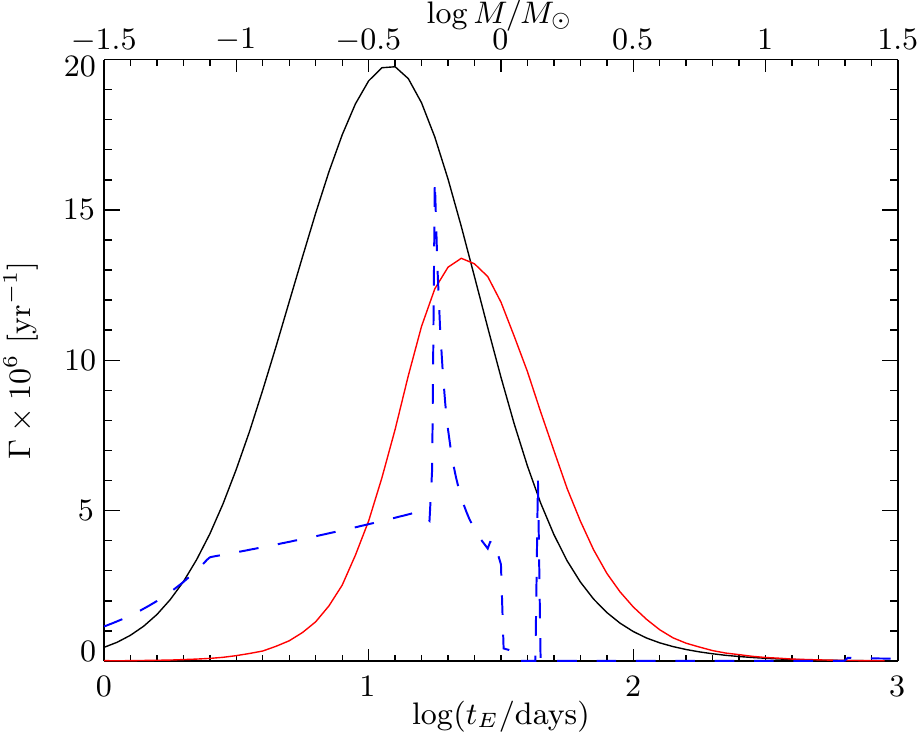}
\caption{The predicted timescale distribution in the direction of Baade's window ($l=1.0^\circ$, $b=-3.9^\circ$). We show as the red line the event rate for a mono-mass $1\msun$ population. The $\sqrt{M_l}$ weighted lens mass distribution is shown as the dashed blue line. When convolved these functions give the black line: the event rate for the assumed mass spectrum of lenses. The rates in this plot, $\Gamma$ are computed per star, per unit $\logten \te$, for stars with $I_0 < 16.7$. Although the timescale distribution is fairly insensitive to magnitude, its normalisation is. The same assumptions on the stellar population as in \autoref{fig:tauvsImag} was used. In the lens mass function we use the assumptions for remnants described in \autoref{sec:pop} and the resultant white dwarfs, neutron stars are visible as the spikes at $0.6\msun$, $1.4\msun$. We use this prescription throughout the paper when timescales are required.  \label{fig:tedist}}
\end{figure}

The event rate is therefore given by \citep{Han:96}
\begin{align}
	\Gamma(\logten \te) &= \int  \gamma(\logten \te') \Phi(\logten M_l) \times  \label{eq:teconv} \\
	&\qquad \qquad \sqrt{M_l} \delta (\logten (\te' \sqrt{M_l}) - \logten \te) d\,\logten \te' d\,\logten M_l \nonumber \\
	&= \int 10^{\frac{1}{2} \logten M_l} \gamma(\logten \te - \frac{1}{2} \logten M_l)  \Phi(\logten M_l) d\,\logten M_l ~. \nonumber
	\label{eq:teconv}
\end{align}
Defining $x=\logten t_E$ and $y=\frac{1}{2} \logten M_l$ then it is clear the event timescale is a convolution
\begin{equation}
\Gamma(x) = 2 \int dy 10^{y} \Phi(2y) \gamma(x - y) ~. \label{eq:teconvobv}
\end{equation}
We can therefore easily compute the event timescale distribution by computing and binning the rate of each particle pair from \autoref{eq:dgammaij} and convolving with the appropriately scaled and weighted mass spectrum using an FFT. 
This provides a recipe for calculation of the event rate and timescale distribution for the $\beta$ parameterisation. To calculate the rate as function of source magnitude, $\Gamma(\log \te,I_0)$, instead a two dimensional convolution is required. In this case we compute for particle pair the weight
\begin{equation}
	d\Gamma_{ij}= \frac{2}{\pi} \frac{\tau_i M_i}{t_{E,ij}} ~.
	\label{eq:dgammaij2}
\end{equation}
This is then two dimensionally binned in equal increments of $\mu$ and $\log \te$. We then convolve this with a kernel of the luminosity function in the distance modulus direction by reference to \autoref{eq:tauvsI0}, and the scaled mass distribution in the timescale direction by reference to \autoref{eq:teconv}. After normalisation with the magnitude distribution, this provides $\Gamma(\log \te,I_0)$, the event rate per star as a function of timescale and magnitude, the primary microlensing observables.

We show in \autoref{fig:bw2deventrate} the event rate $\Gamma(\log \te,I_0)$ in the direction of Baade's window. While the event rate is a strong function of source magnitude, $I_0$, the timescale distribution is not. Only in the tilt of the red clump region is there a visible change.

We show in \autoref{fig:tedist} the event timescale distribution. This shows how this distribution arises through convolution of the mono-mass timescale distribution of the dynamical model, with the distribution of lens masses. The resultant timescale distribution is a useful probe of the mass distribution lenses in particular which we briefly consider in \autoref{sec:tedistdiscuss}. 

\begin{figure*}
\includegraphics[width=\linewidth]{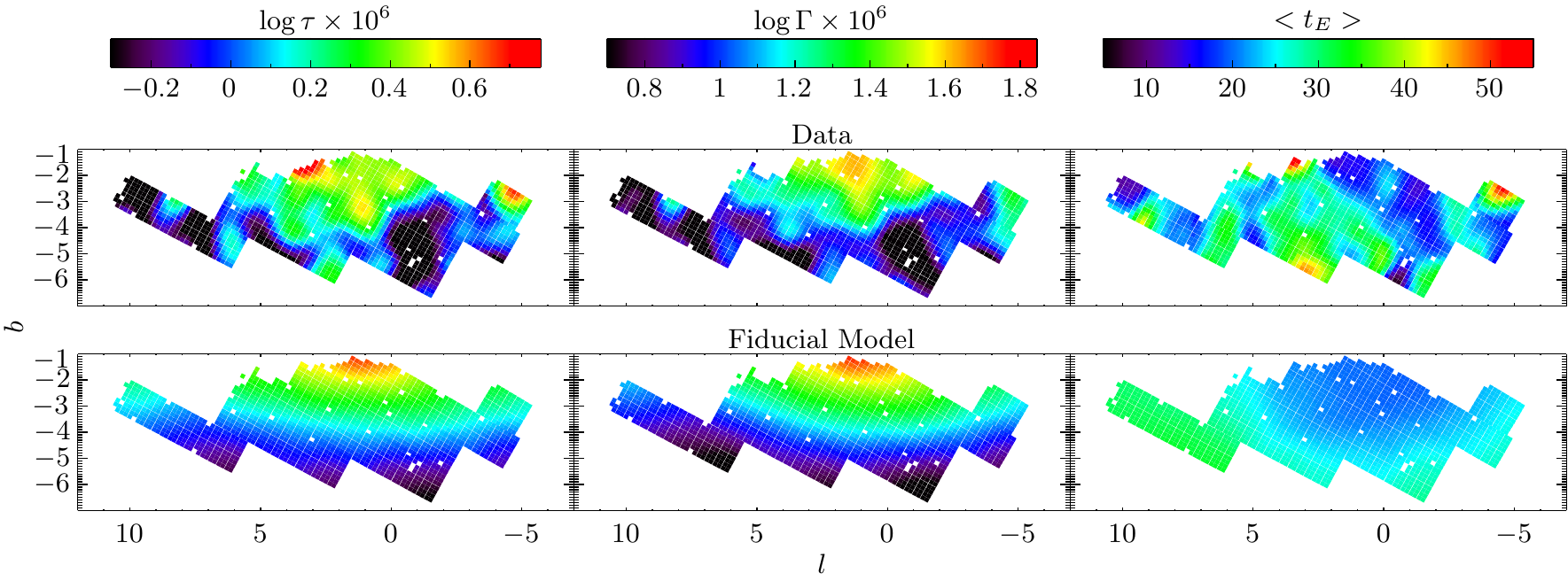}
\caption{Comparison of the predicted optical depth, $\tau_{200}$ of our fiducial model  (bulge model M90, together with a disk of scale length $R_d=2.6\kpc$ and an inner scale height of $H_{4.5}=0.18\kpc$) compared to the MOA-II data corrected by the revision in \citep{Sumi:16}. The data and the model are averaged in the same manner: an Gaussian average with $\sigma=0.4\deg$ of fields within $1\deg$. We show from left to right comparison of the optical depth, the event rate per star, and the mean timescale $\langle \te \rangle$ in days. The optical depth and event rate maps show good qualitative agreement. The mean timescale map has large observational error due to the large statistical variance of timescale caused by the rare very long timescale events. The timescale map is better constrained by the larger sample of \citet[][\eg Figures 12, 14 and 16]{Wyrzykowski:15} and we discuss its distribution in \autoref{sec:tedistdiscuss}. \label{fig:sumifidmaps} }
\end{figure*}

\section{Constraints from MOA-II Microlensing Data}
\label{sec:data}

In this section we compare our Milky Way models to the sample of microlensing data from the MOA-II survey \citep{Sumi:13}. Throughout we use the revised microlensing optical depth and rate data corrected for the effective number of monitored stars according to \citet{Sumi:16}.

Both the original \citep{Sumi:13} and the revised MOA-II data \citep{Sumi:16} split their sample of microlensing events an all-source sample, defined as sources with $I<20\mags$, and an red clump giant (RCG) sample restricted to $I<17.5\mags$. The selection criteria of MOA-II was optimised to detect short period events, however bright long timescale events are not efficiently detected: the longest period event in the RCG sample is only 55 days. Because these events make a significant contribution to the optical depth in the RCG sample we therefore consider only the all-source sample where the issue is not significant, as do \citet{Sumi:16}. The all-source sample contains events with $t<200\,{\rm days}$ and therefore throughout we compute and compare the data with the same cut. In \autoref{sec:eros} we compare to the EROS-II microlensing optical depth data measured by \citep{Hamadache:06} of bright stars around the red clump from to show the results are consistent.  

\subsection{Microlensing Maps}
\label{sec:maps}

Here we compute maps of the optical depth, timescale and event rate of our fiducial model to compare to the maps in \cite{Sumi:16}.  

We use a simple exponential model for the extinction:
\begin{equation}
	\frac{dA_I}{ds} = a_I \times \exp ( -|z|/z_0 )
\end{equation}
with coefficients $a_I=0.5\mags\kpc^{-1}$ and $z_0=160\pc$ \citep{Nataf:13}. We have tried other values for these coefficients, and found that with the magnitude cuts selected by \citet{Sumi:13} the results are generally insensitive to these choices. For the latitudes probed by current microlensing experiments the source stars lie at $z \gg z_0$ and therefore extinction is largely a foreground screen. The changes due to extinction therefore arise mostly because extinction reduces the unextincted magnitude limit (see \eg \autoref{fig:tauvsImag}). This therefore changes the optical depth by  altering the range of magnitudes over which the measured optical depth is an average. Extinction has a much greater effect on the rate per unit area, and we therefore choose not to model this quantity.

We compute the microlensing properties of the models in each of the 1536 subfields of MOA-II using the procedure described in \autoref{sec:tetheory} to predict $\Gamma(\log \te, I)$. We then average the optical depth, weighted by the model magnitude distribution, over the range $14<I<20$ to represent the MOA-II magnitude range. 

We exclude events with $\te>200\days$ when computing the optical depth \ie we calculate $\tau_{200}$. Both the event rate and mean event timescale have no exclusion of $\te>200\days$ events. Maps comparing the fiducial model with the data of the resultant optical depth $\tau_{200}$, the event rate $\Gamma$, and the mean event timescale $\langle \te \rangle$ are shown in \autoref{fig:sumifidmaps}.

\subsection{Minor Axis Profile}
\label{sec:minoraxis}

\begin{figure}
\includegraphics[width=0.9\linewidth]{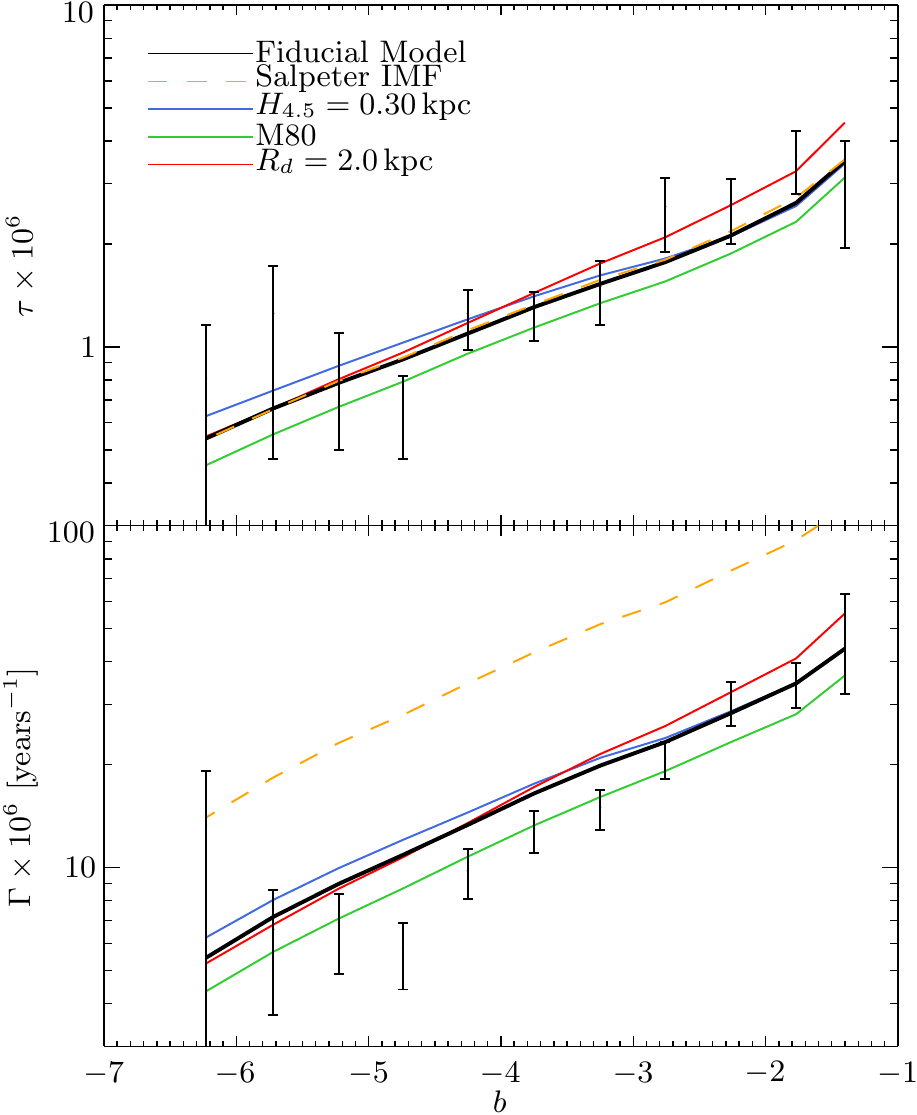}
\caption{Comparison of the predicted microlensing properties for $|l|<5\deg$ with the revised MOA-II data. The upper panel shows the optical depth, and the lower the microlensing rate per star. The fiducial model, summarised in \autoref{tab:model}, is the solid black line. The coloured lines correspond to changing one feature of the fiducial model: blue to a constant $H=0.3\kpc$ scale height disk, green to bulge mode M80 with a lower stellar mass in the bulge, and red to a $R_d=2\kpc$ disk scale length. The orange line corresponds to the effect of changing to a \citet[][]{salpeter:55} IMF. The impact of the uncertain IMF on the rate leads us to constrain the models using the optical depth.  \label{fig:minor_sumicomp}}
\end{figure}

The maps in \autoref{fig:sumifidmaps} show qualitatively good agreement. Here we use the minor axis profile of the microlensing data in the range $|l|<5\deg$ to make \emph{quantitative} comparison to Galactic models and hence determine which models are allowed by the data. The revised MOA-II data is taken from Table 2 of \citet{Sumi:16}. We take the maps computed from the models in the previous section and bin the $|l|<5\deg$ microlensing properties in the same manner as the data. The result for the fiducial model compared to the data is shown as the black solid line in \autoref{fig:minor_sumicomp}. We show both the optical depth and rate predictions. The fiducial model matches the optical depth well with a $\chi^2$ of $7.4$ over the 11 data points \ie 0.7 per data point. 

We also show the effect of variations to our fiducial parameters in \autoref{fig:minor_sumicomp}.  In particular we show the effect of changing the IMF from our fiducial \citet{Kroupa:01} IMF, to a \citet{salpeter:55} IMF. This effects the event rate significantly, but hardly changes the optical depth. This is because the IMF directly alters the event timescale distribution and therefore directly alters the rates. However the IMF only enters the $\tau_{200}$ optical depth through the timescale distribution altering the contribution of events beyond the 200 day cutoff. For this reason we focus on constraints provided by the optical depth alone, and discuss the timescale distribution separately in \autoref{sec:tedistdiscuss}. As shown in that section our fiducial \citet{Kroupa:01} IMF has a slightly shorter timescale distribution than the observed. Using an IMF with less low mass stars matches the timescale distribution better, and would therefore also match better the rate profile data in the lower panel of \autoref{fig:minor_sumicomp}.

\subsection{Constraining the Model}
\label{sec:Rdconstaint}

\begin{figure}
\centering
\includegraphics[width=0.9\linewidth]{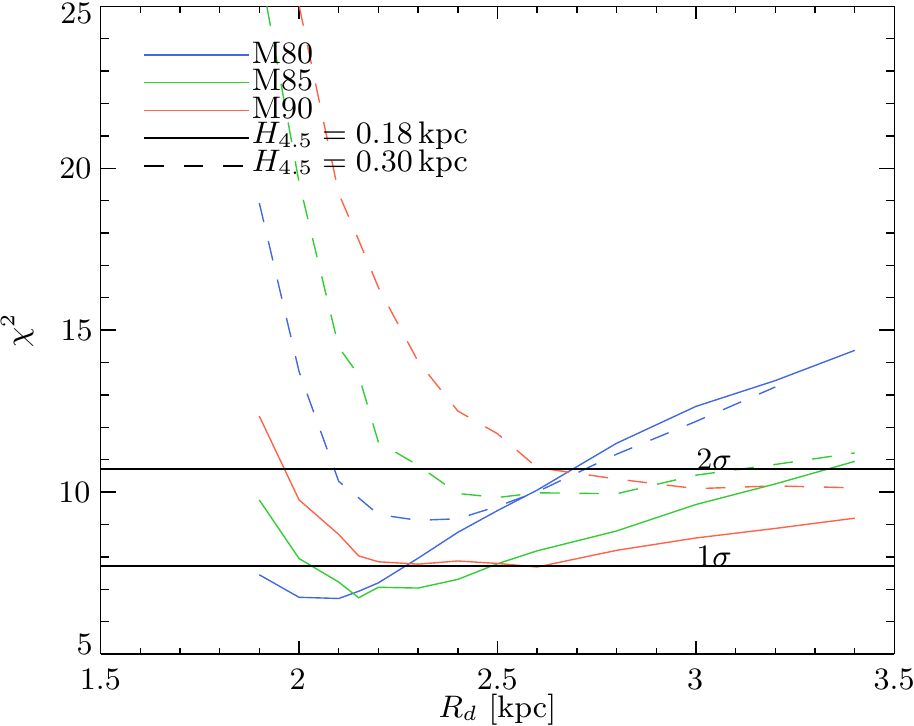}
\caption{Allowed disk scale lengths as a function of $R_d$ for the revised MOA-II optical depth data. $\chi^2$ is calculated from the difference between model and data as in \autoref{fig:minor_sumicomp}. Different colours correspond to bulge models with different stellar masses M90 having the highest stellar mass and M80 the lowest. The dashed lines correspond to a constant $0.3\kpc$ scale height disk and the solid lines to a flared disk with inner scale height $H_{4.5}=0.30\kpc$. \label{fig:Rdconstaint}}
\end{figure}

In \autoref{fig:Rdconstaint} we show the $\chi^2$ between the data and model, computed using the data in \autoref{fig:minor_sumicomp}, for each bulge model as a function of disk scale length, $R_d$. The different lines correspond to variations of the bulge dark matter fraction and disk scale length. We consider only $R_d \ge 1.9\kpc$ since shorter disk scale lengths produce a baryonic rotation clearly larger than observed, with a peak $V_c>240\kms$ as computed in \autoref{sec:rotcurve}. 

We consider the bulge dynamical models with different bulge dark matter fractions (M80-M90) and consider both  the flared disk with our fiducial inner scale height of $H_{4.5}=0.18\kpc$ and a constant disk scale height of $H_{4.5}=H_\odot=0.3\kpc$. There is some degeneracy between disk scale length and bulge model: The bulge model with lower stellar matter, M80, requires short disk scale lengths of $(2.1 \pm 0.2)\kpc$  optical depths (at $1\sigma$). However model M85 with a larger stellar fraction would require $(2.3 \pm 0.2)\kpc$, and model M90 with the highest stellar mass fraction prefers even longer disks. This degeneracy occurs because the optical depth, \autoref{eq:tau}, measures the density of stellar matter towards the bulge, however for the same optical depth the lenses can be placed either in the bulge region or the foreground disk.  We therefore consider the different bulge models and disk models together in the following sections. 

The flared disks with inner scale height of $H_{4.5}=0.18\kpc$ are  preferred at $>1\sigma$ over the constant scale height $H_{4.5}=0.3\kpc$ disks. This is because the constant scale height disks produce slightly flatter optical depth vs. latitude profiles than the data. For the same bulge model the constant disk scale height models require longer disk scale lengths. This is because most of the optical depth points lie higher than one disk scale height from the Galactic plane. Therefore a larger inner disk scale height places more mass at latitudes where it contributes to the optical depth and reduces the need for additional stellar mass in the disk. 

\section{Comparison to EROS-II Microlensing Optical Depths}
\label{sec:eros}

In this section we consider if the models consistent with the revised MOA-II optical depth data are also consistent with the EROS-II optical depths. \citet{Hamadache:06} measured the EROS-II optical depth from a sample of bright resolved events. Only events magnifying stars in a selection box centred on the red clump in color-magnitude were considered. We replicate this in the models by computing the magnitude distribution in each of the 66 EROS-II bulge fields and in these fitting for the magnitude of the simulated red clump. We then compute the optical depth of stars within $1\, {\rm mag}$ of this to simulate the selection box in each field, and combine the optical depths across fields to simulate the optical depth vs latitude profile measured by \citet{Hamadache:06}.

\begin{figure}
\centering
\includegraphics[width=0.9\linewidth]{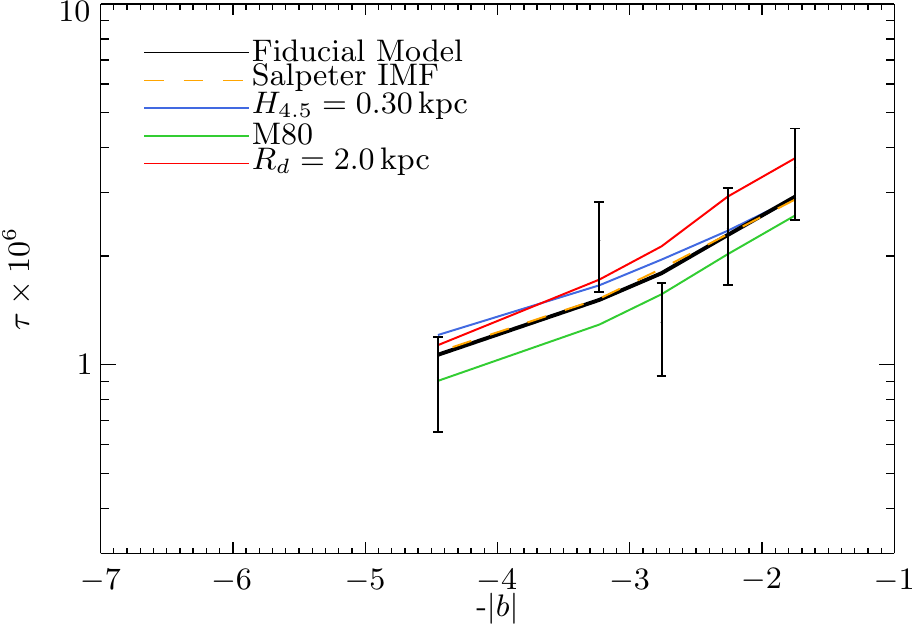}
\caption{Comparison of the predicted microlensing optical depth for $|l|<5\deg$ compared to the EROS-II data  \citep{Hamadache:06}. The fiducial model is the black line and the other lines correspond to the same models as in \autoref{fig:minor_sumicomp}. \label{fig:minor_eroscomp}}
\end{figure}

\begin{figure}
\centering
\includegraphics[width=0.9\linewidth]{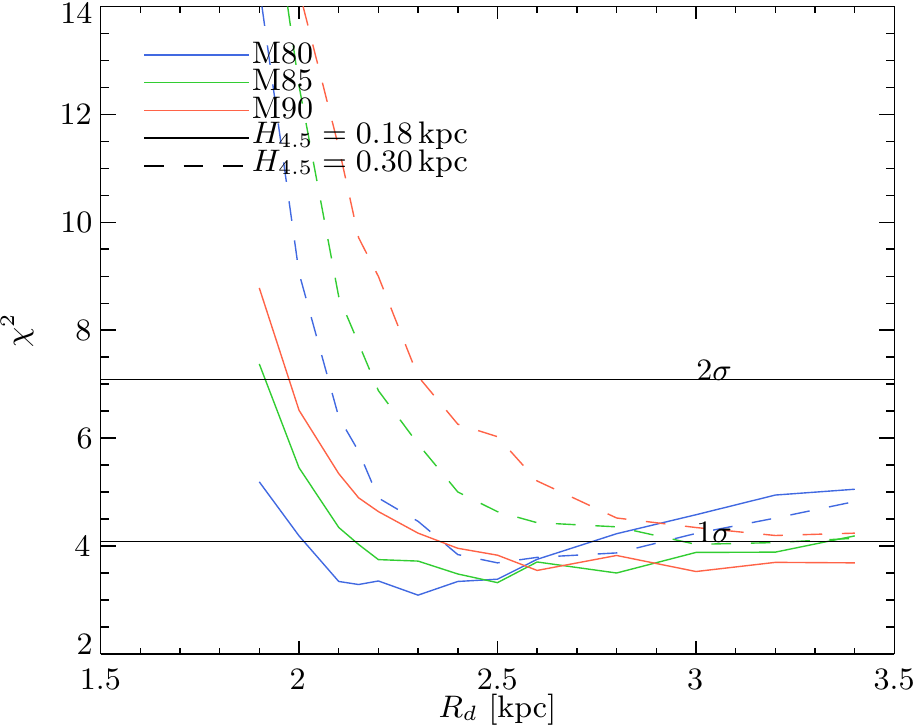}
\caption{The $\chi^2$ of the Galactic models as a function of $R_d$ for \citet{Hamadache:06} optical depth data. $\chi^2$ is calculated from data points as in the examples in \autoref{fig:minor_eroscomp}. The lines correspond to the same models as in \autoref{fig:Rdconstaint}.  \label{fig:erosRdconstaint}}
\end{figure}

The resultant profiles are shown compared to the data in \autoref{fig:minor_eroscomp}. The same fiducial model fits the EROS-II data with a $\chi^2=3.5$ across 5 data points, or 0.7 per data point. In \autoref{fig:erosRdconstaint} we show the models that are consistent with the data. Comparison with \autoref{fig:Rdconstaint} shows that the two data sets constrain the models similarly, and that models consistent with the MOA-II data are also consistent with the EROS-II data. The constraints provided by the EROS-II data are however weaker. This is a result of the smaller number of events considered: 120 in EROS-II vs 427 in MOA-II.

Initial microlensing studies yielded optical depths that were larger than theoretical predictions \citep{Udalski:94,Alcock:97}, and even theoretical bounds \citep{Binney:00}. Some these high optical depths resulted from an inadequate treatment of blending of faint stars, a problem that was not serious in the samples of brighter RCG stars \citep{Popowski:05}. The situation improved with DIA analysis and better treatment of blended stars. It is reassuring then that the models which match the revised MOA-II all-source data, also match the optical depths of the bright resolved RCG sample from \citet{Hamadache:06}, albeit with larger statistical errors associated with their smaller sample.

\section{Constraining the Galactic Rotation Curve with Microlensing Data}
\label{sec:rotcurve}

\begin{figure}
\includegraphics[width=\linewidth]{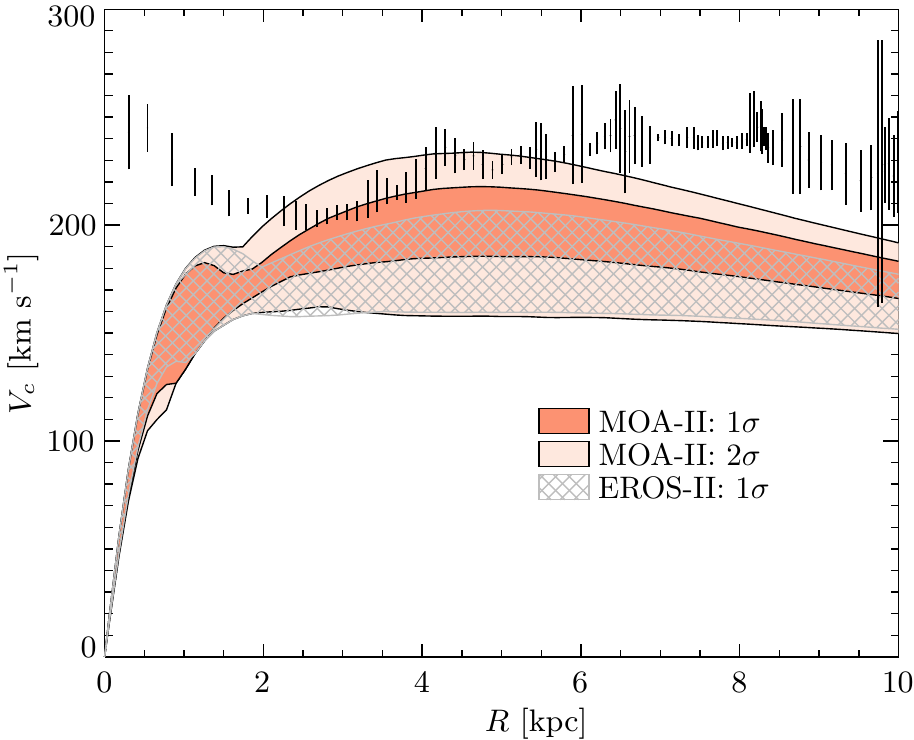}
\caption{Bands are the range of baryonic rotation curves of the models consistent with the revised MOA-II data \citep{Sumi:16} at 1 (red) and $2\sigma$ (pink). The baryonic rotation curves of the models consistent with the EROS-II data \citep{Hamadache:06} at $1\sigma$ level span the cross hatched area. The models plotted here have a range of bulge dark matter fraction (M80-M90), disk scale length ($R_d$) and inner disk scale height ($H_{4.5}$) but are all consistent with the microlensing optical depths. Data is taken from the compilation of gas kinematics by \citet{Sofue:09} scaled to $R_0=8.3\kpc$ \citep{Sotiris:14} and $V_0=238\kms$ \citep{Schonrich:12} (see \citealt{Ortwin:araa} for a fuller discussion of these values).   \label{fig:rotcurve}}
\end{figure}

Each of the models consistent with the microlensing data, as computed in \autoref{sec:Rdconstaint}, has an associated stellar mass and, thus, stellar rotation curve.  We include in addition a gas component of surface mass $13\,M_\odot\pc^{-2}$ with scale height 130\pc and scale length $2R_d$ \citep{Bovy:13}. We plot in \autoref{fig:rotcurve} the resultant rotation curves consistent with the revised MOA-II data and those consistent with the EROS-II data \citep[\cf Fig 16 of][]{Ortwin:araa}. We also plot the circular velocity taken from the compilation of gas dynamics of \citet{Sofue:09} scaled to $V_0=238\kms$ \citep{Schonrich:12,Reid:14} and $R_0=8.3\kpc$ \citep{Sotiris:14}.

When performing disk-halo decompositions in external disk galaxies there is a well known stellar mass-to-light vs. dark matter degeneracy \cite[\eg][]{vanAlbada:85,Courteau:14}. The maximal disk hypothesis breaks this by assuming the disk contributes the maximum level possible by the data. Under this assumption typical disk contributions to the circular velocity at $R=2.2R_d$ in external disk galaxies are $f_v=(0.85\pm0.10)$ \citep{Sackett:97}. The point $R=2.2R_d$ is chosen as the comparison point since this is the location of the maximum of the rotation curve of a thin exponential disk.

The specific value used as the boundary between maximal and sub-maximal disks is arbitrary. Maximal disks necessarily contribute less than $f_v=1$ because the halo is significant at large radii, and its density cannot reasonably decrease inwards. However the lower end of $f_v=(0.85\pm0.10)$ given by \citet{Sackett:97} for maximal disk fits results from bulge-disk decompositions where the bulge is significant and not included in \fv~. Motivated by this, and following \citet{Courteau:14}, we consider $\fv=0.85$ as the boundary for maximal disk models.

Microlensing of Milky Way bulge stars offers a unique way to break the baryonic to dark matter degeneracy, and therefore estimate dark matter fractions and test the maximal disk hypothesis. However making a similar comparison for the Milky Way is complicated by the Galaxy not being a pure disk.  Recently it has been realised that the bulge of the Milky Way is consistent with a minimal contribution from a classical bulge \citep[\eg $<8\%$ of the disk mass by][]{Shen:10}. Instead the  the bulge is a box/peanut bulge which formed via secular evolution from the disk. We therefore consider instead the fractional baryonic contribution at the peak of the baryonic rotation curve, since this seems most analogous to disk-halo decompositions in external galaxies \citep[as also advocated by][for external galaxies]{Courteau:14} and use the same boundary $\fv=0.85$ for maximality. The total circular velocity used for comparison is the weighted average of points within 500\pc of this computed similarly to \autoref{fig:rotcurve}.

For each model admitted by the data we compute this maximality: the fractional baryonic \fv. From the extremes of the models consistent with the revised MOA-II data at $1\sigma$ we find a maximality  \onesigbounds. Performing the same process with the models consistent with the EROS-II data gives  \onesigboundseros~at $1\sigma$, the wider range of allowed curves resulting from the smaller number of microlensing events. These levels of baryonic contribution therefore place the Milky Way on the boundary between maximal disk fits and sub-maximal disks. 

At $2\sigma$ a wider range of models are allowed. No useful upper limit on the maximality can be derived because very short disk scale lengths are allowed which would result in a baryonic contribution that even exceeds the total rotation curve. Instead only a lower limit of $>0.75$ can be given. However the models with high stellar to dark matter fractions in the bulge admit very long disk scale lengths since, for these bulges at $2\sigma$, the MOA-II optical depths can be reproduced almost by the bulge alone. This can result in baryonic rotation curves whose peak lies very close to the Galactic centre at $<2\kpc$, and which then slowly decline outwards. Computing the maximality at this position for these models is suboptimal for two reasons
\begin{inparaenum} 
\item the circular velocity from \citet{Sofue:09} here is very uncertain here because the bar influences the gas orbits making them highly non-circular
\item the maximality is useful as a number that encodes the interplay between baryonic and dark matter in galaxies, however making the comparison at such a small fraction of the disk scale-length reduces its utility.
\end{inparaenum}
We therefore also consider a second Galactocentric radius for comparison. The models consistent with the data at $1\sigma$ have their baryonic peaks at $(4-5)\kpc$ from the Galactic centre. If we compute the maximality at $4\kpc$ the maximality constraint would be slightly lower: $\twosigmaxfrac$ at $2\sigma$ and this is the value we consider more robust. From EROS-II no useful constraints can be extracted at $2\sigma$.

\section{Timescale Distribution}
\label{sec:tedistdiscuss}

\begin{figure}
\includegraphics[width=\linewidth]{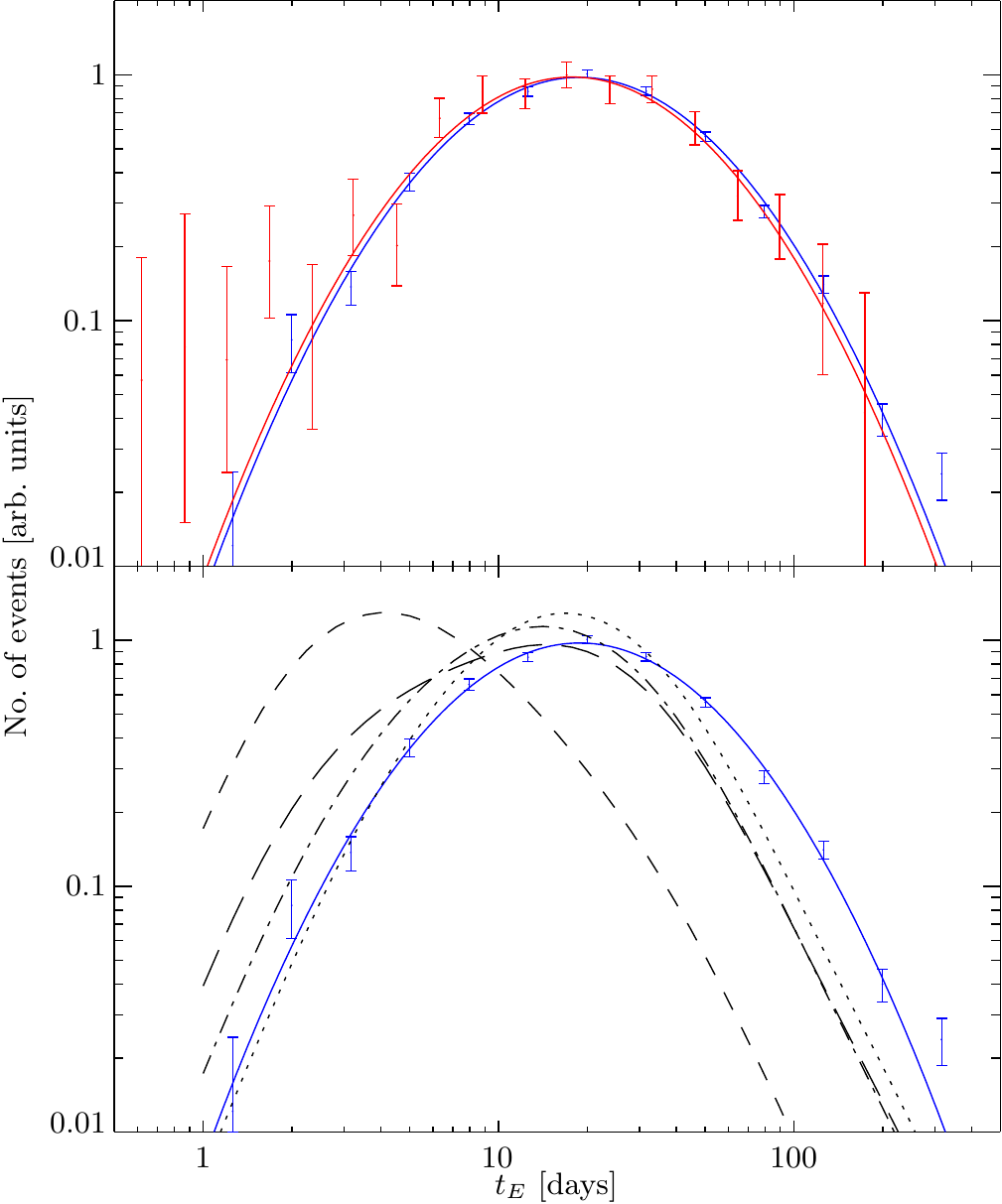}
\caption{In the above panel we show the efficiency corrected timescale distribution of the OGLE-III survey (blue) compared to the MOA-II sample (red). We also show the best fitting log-normal distributions to the surveys as the curves. In the lower panel we compare the efficiency corrected OGLE-III timescale distribution to our fiducial model with different IMFs: A \citet{salpeter:55} IMF (short dashed line), a \citet{Zoccali:00} IMF (long dashes), a \citet{Kroupa:01} IMF (dash-dot) and the log-normal IMF of \citet{Calamida:15} (dotted). \label{fig:tevssurvey}}
\end{figure}

Although the focus of this work are the constraints that surveys of microlensing in the bulge place on Galactic structure, which is best revealed by the optical depth, we briefly consider the timescale distribution in this section. While the optical depth has not yet been computed from the OGLE-III survey, the timescale distribution has by \citet{Wyrzykowski:15}. In the upper panel of \autoref{fig:tevssurvey} we show a comparison of this with the timescale distribution measured by MOA-II. 

The two samples agree in their timescale distribution remarkably well. Fitting a log-normal distribution to each we find the mean log timescales are $<\log \te>=(1.275\pm0.008)$ and $(1.25\pm0.03)$ for the OGLE-III survey and MOA-II all source sample respectively, in statistical agreement. Likewise the standard deviation of the log timescales are $\sigma(\log \te)=(0.409\pm0.006)$ and $(0.41\pm0.02)$, also in agreement.  

Note that a log-normal distribution does not capture the short timescale events present in the MOA-II sample. These are presumed to arise from microlensing by planetary mass objects \citep{Sumi:11}. It is also not expected to capture the asymptotic distribution of short or long timescale events \citep{Mao:96} but, given the current data, is still a useful parameterisation. This is particularly true since it provides the first two moments of the log timescale distribution. The log timescale distribution is a convolution of the two PDFs in \autoref{eq:teconvobv}, resulting from the dynamical model, and the lens mass distribution. Because of this the means and variances add (and indeed all cumulants) to give the mean and variance of the log timescale distribution. The mean log timescale and its variance are therefore straightforward to interpret and consider the effects of other mass distributions or dynamical models. 

In the lower panel of \autoref{fig:tevssurvey} we show the timescale distribution predicted from our model for four different IMFs: those from \citet{salpeter:55} (a single power law with slope $\alpha=2.3$), a \citet{Zoccali:00} IMF ($\alpha=1.3$ for $M<1\msun$), a \citet{Kroupa:01} IMF, and log-normal IMF fitted for the bulge by \citet{Calamida:15} (log-normal with $M_c=0.25\msun$, $\sigma=0.5$). All IMFs use lower and upper mass limits of $0.01\msun$ and $100\msun$ respectively, aside from the \citet{salpeter:55} IMF which uses a lower limit of $0.1\msun$. For all we use the remnant prescriptions of \citet{Maraston:98} above the turnoff of a $10\Gyr$ population as described in \autoref{sec:pop}.

Given that we have not tailored the model to the timescale distribution it is reassuring that the \citet{Kroupa:01} and particularly the \citet{Calamida:15} IMFs together with the dynamical model reproduce the timescale distribution fairly well. There is a discrepancy at long timescales which the model under predicts. As a result, with the \citet{Calamida:15} IMF the model predicts $<\log \te>=1.21$ slightly less than the observed $<\log \te>=(1.275 \pm 0.008)$. The most uncertain region of the mass function is the brown dwarf region. That the short timescale distribution agrees with the OGLE-III data for the \citet{Calamida:15} log-normal IMF suggests that our model requires a low number of brown dwarfs to be present in the bulge compared to rising power law IMFs like the \citet{Kroupa:01} IMF, similar to local estimates \citep[\eg][]{Andersen:08}.

More detailed comparison is beyond the scope of this work and would require consideration of variations of the kinematical model, particularly outside the central region where the N-body model was constrained by \citet{Portail:15}. We are presently constructing of made-to-measure models of the entire inner Galaxy, including the inner disk, and  we will model the timescale distribution and place more robust constraints on the IMF when that is complete. We emphasise that the results outside this section are based on the optical depth, which is insensitive to the timescale distribution. This is demonstrated in \autoref{sec:minoraxis} by the insensitivity of the optical depth to the much shorter timescales produced by the \citet{salpeter:55} IMF.

\section{Discussion}
\label{sec:discuss}

\subsection{A high baryonic fraction in the inner Milky Way}

The quantity best constrained by the microlensing optical depth is the stellar mass per unit solid angle towards the bulge and inner Galaxy. We find that there is a degeneracy in where that mass is placed between the bulge and the foreground disk: models with lower stellar fractions in the bulge require more prominent foreground disks and vice versa. 

The scale length of the Milky Way disk is highly uncertain. Earlier estimates using optical wavelengths tended to favour larger scale lengths \citep[\eg 3.5\kpc by][]{Bahcall:80}, while more recent measurements in the NIR and large photometric parallax surveys have favoured lower measurements typically in the range 2-3\kpc ( \eg \citealt{Binney:97,Bissantz:02,Robin:03,Juric:08,Bovy:13}; see \citealt{Ortwin:araa} for a fuller discussion).

Despite this uncertainty, in all the models which reproduce the revised MOA-II optical depths, the baryonic contribution to the rotation curve at its peak is high: \onesigbounds~at $1\sigma$ and $\twosigmaxfrac$ at $2\sigma$. These are consistent with the constraints from EROS-II of \onesigboundseros. As discussed in \autoref{sec:rotcurve}, we consider maximal disks to have $\fv > 0.85$ \citep{Courteau:14}. These levels of baryonic contribution therefore place the Milky Way on the boundary between maximal disk fits and sub-maximal disks . 

The DiskMass survey has measured the vertical velocity dispersions of nearly face on disk galaxies. Together with the statistically determined scale height from a sample of side on galaxies, this directly measures the disk mass under the assumption of a locally isothermal disk. They find that their sample is generally sub-maximal with a fractional baryonic contribution at $2.2R_d$ of $0.57\pm0.07$ \citep{Martinsson:13}. However the measured light weighted stellar kinematics will be biased towards younger stars as opposed to old dynamically relaxed populations. Measuring the vertical kinematics of the Milky Way locally suggests that this correction could be a factor of $\sim 2$ \citep{Aniyan:16}. In addition the resultant IR mass-to-light of the disk mass survey is also a factor of $\sim 2$ smaller than that estimated through other methods \citep{McGaugh:15}. The derived maximal disk in this work would therefore not necessarily make the Milky Way unusual in the context of external disk galaxies. In addition in external barred galaxies a further possibility exists to break the baryonic to dark matter degeneracy. The stellar distribution can be determined from the light, while the non-circular gas motions constrain the shape of the effective potential, and therefore the more spherical dark matter, in the bar. Using this method several studies have found maximal or near maximal disks in the inner regions of barred galaxies \citep{Weiner:01,Weiner:04,Perez:04,Sanchez:08}.

\subsection{Consequences for the Milky Way's dark halo}

\begin{figure}
\includegraphics[width=\linewidth]{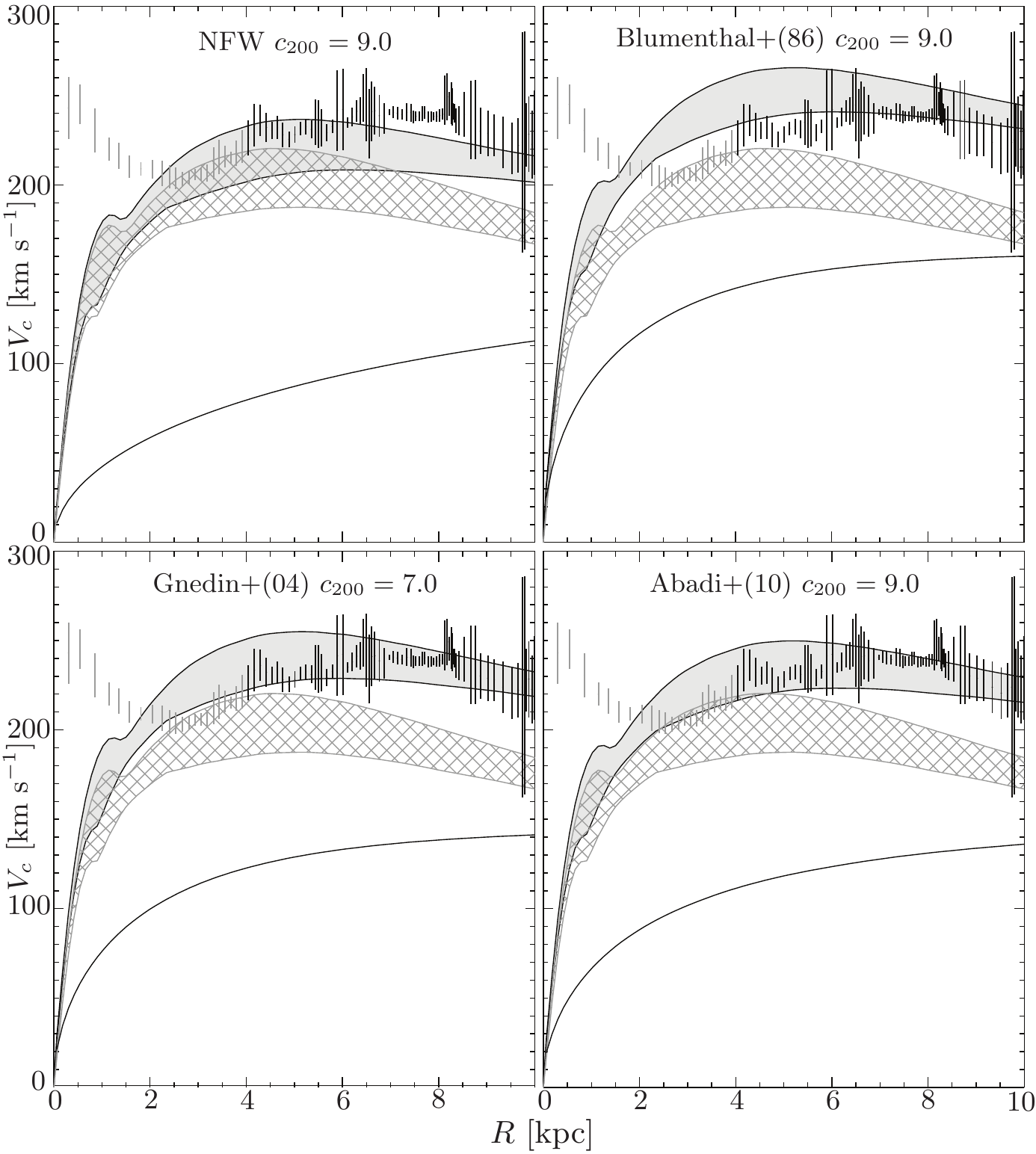}
\caption{We show the range of baryonic rotation curves consistent with the revised MOA-II data at the $1\sigma$ level together with a selection of possible dark matter haloes. The baryonic rotation curves are shown as the cross hatched area, the dark matter profiles as the black lines, and the resultant range of profiles of the total rotation curve in grey. The top left panel shows an NFW profile with concentration $c_{200}=9.0$. The upper right an NFW halo, also with $c_{200}=9.0$,  adiabatically contracted through the prescription \citet{Blumenthal:86}. The lower left panel shows the   contraction suggested by \citet{Gnedin:04} for a smaller but still cosmologically possible initial $c_{200}=7.0$, and the lower right the contraction fitted by \citet{Abadi:10} for $c_{200}=9.0$. \label{fig:rotadiabatic} }
\end{figure}

The high baryonic contribution to the rotation curve in the inner Galaxy seen in \autoref{fig:rotadiabatic}, and the resultant high levels of disk maximality, are not however inconsistent with the current understanding of the $\Lambda$CDM paradigm. An NFW dark matter halo with a Milky Way halo mass $M_{\rm 200}=1.1 \times 10^{12} M_\odot$ and radius $R_{\rm 200}=270\kpc$ \citep{Ortwin:araa} and a concentration of $c_{\rm 200}=9.0$ motivated by cosmological simulations (\citealt{Correa:15c} with Planck 2015 cosmology \citealt{PlanckCosmo:15}) gives the profile shown in the top left panel of \autoref{fig:rotadiabatic}. Adding this to the range of baryonic contributions allowed by the revised MOA-II data at $1\sigma$ gives rotation curves that remain below the total rotation curve. 

\begin{figure*}
\includegraphics[width=0.7\linewidth]{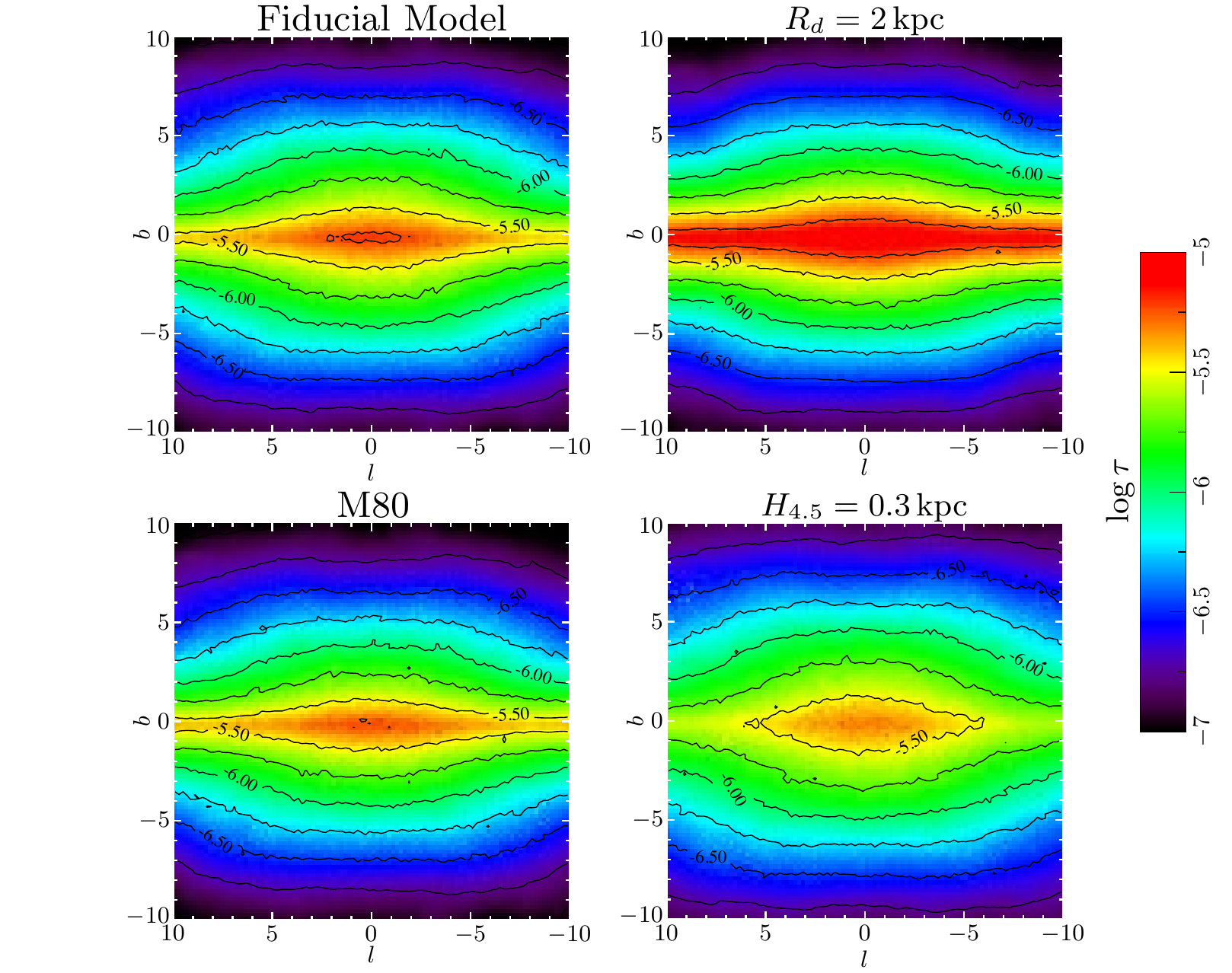}
\caption{The microlensing optical depth of the fiducial model across the entire Milky Way bulge (top left), compared to the same variants considered earlier in the paper (\eg in \autoref{fig:minor_sumicomp}). We show the effect of reducing the disk scale length to $R_d=2\kpc$ from $R_d=2.6\kpc$ in the top right, using a bulge model with a higher dark matter fraction and lower stellar mass in the bulge (M80) in the lower left, and a model with a constant scale height of $H_{4.5}=0.3\kpc$ in the lower right. All maps are the average optical depth over stars with $14<I_0<19$. \label{fig:fullmaps}}
\end{figure*}

Adiabatically contracted versions of this halo using the prescription of \citet{Blumenthal:86} are generally not consistent with the revised MOA-II data at the $1\sigma$ level. In this prescription  the adiabatic invariant is $r M(<r)$ which results in rotation curves that overshoot that observed in the inner Galaxy (top right panel \autoref{fig:rotadiabatic}). This is true even for lower concentrations as small as $c_{\rm 200}=7.0$, which is the expected cosmological halo-to-halo scatter at $1\sigma$ level \citep{Maccio:08}. It is also true for the contraction detirmined by \citet{Piffl:15} since it results in more contracted haloes than \citet{Blumenthal:86}.  \citet{Binney:15} similarly concluded that halos with this level of contraction, in conjunction with the local dark matter density, could not both be consistent with the Milky Way's rotation curve and earlier microlensing optical depths.

However other milder contraction prescriptions are consistent with the revised MOA-II data at the $1\sigma$ level. The prescription of \citet{Gnedin:04} is consistent only if the initial halo has lower concentrations ($c_{\rm 200}=7.0$ is shown in the lower left panel of \autoref{fig:rotadiabatic}). The prescription of \citet{Abadi:10} results in less contracted halos, and as a result is consistent with an initial $c_{\rm 200}=9.0$ (lower right panel of \autoref{fig:rotadiabatic}). All haloes were contracted using a baryonic exponential scale length of $2.6\kpc$ and contraction of the \citet{Gnedin:04} and \citet{Abadi:10} halos was implemented through the $\Gamma$ prescription of \citet{Dutton:07,Dutton:11} with $\Gamma=0.8$ and $0.4$ respectively. All contraction prescriptions can be consistent with the range of rotation curves allowed by the MOA-II data at the $2\sigma$ level, and the EROS-II data.

It has been shown that strong feedback could alter the central profiles of Milky Way sized haloes \citep{Maccio:11}. The results here suggest that it is not \emph{required} in the centre of Milky Way sized galaxy haloes in a similar way to dwarf galaxies. The microlensing results however do not rule it out, for example the feedback in \citet{Chan:15} for $10^{12} M_\odot$ mass haloes results in central profiles similar to the pre-contracted NFW haloes which would still be consistent. We would expect our results to also be consistent with other recent simulations \citep[\eg][]{Marinacci:14,Schaller:15} which generally produce only mildly contracted haloes.

We conclude that the levels of baryonic contribution towards the inner Galaxy, \fv,  are high, but not inconsistent with the estimates of contribution of CDM haloes in the inner region of disk galaxies, provided the level of  baryonic contraction is not too extreme. 

There is however possible tension between the low baryonic fractions in the inner galaxy measured here, with recent estimates of the local dark matter density. \citet{Piffl:14} measured the local dark matter density to be $0.0126 q^{-0.89} \msun \pc^{-3} \pm 15\%$  in agreement with  $0.014 \msun \pc^{-3}$ measured by \citet{Bienayme:14}. These more recent measurements utilising the exquisite RAVE data are higher than most over that past decade \citep[see][and references therein]{Read:14}. The haloes shown in \autoref{fig:rotadiabatic} however have local dark matter densities in the range $0.005-0.008 \msun pc^{-3}$.

The NFW profile considered by \citet{Piffl:14} would be consistent with the range of models considered here and would reproduce also local dark matter density. However this model had an uncomfortably high concentration of $c=20$, significantly higher than expected in dark matter only simulations for Milky Way sized halos. Instead it seems more natural that a less concentrated halo experienced a mild degree of contraction due to growth of the baryonic disk, and this increased the dark matter density at $\sim R_0$. Some tension remains however between the high baryonic fraction in the inner Galaxy measured here, the local dark matter density, and even mild adiabatic contraction prescriptions. Contracting a halo with $c_{200}=15.$, towards the upper end of the cosmological values, with the prescription of \citet{Abadi:10} results in rotation curves just consistent with the microlensing results here. This however gives a solar dark matter density of $0.01 \msun pc^{-3}$, still somewhat lower than the estimates with RAVE. We therefore encourage direct comparison between the high baryonic fraction in the inner Galaxy required by the microlensing here, the local dark matter density, and recent simulations of Milky Way like galaxies in the $\Lambda$CDM paradigm such as \citet{Marinacci:14}. 

\subsection{Microlensing as a tool for Galactic structure}

The measured optical depths close to the plane are particularly important in driving the need for high stellar contributions in the inner Galaxy and low dark matter fractions, because the majority of the stellar mass lies here. It is therefore important to confirm the optical depth measurements in this region. 
We anticipate that the constraints in this work could be verified and improved on by the larger sample of 3560 events, and the better detection efficiency of longer events, provided by the OGLE-III sample  \citep{Wyrzykowski:15}. Of these 2047 lie inside $|b|<3\deg$ and 546 inside $|b|<2\deg$, therefore the optical depths measured by MOA-II in this region can be verified. The optical depth and event rates require detailed field-by-field efficiency calculations  that are not yet complete, but we encourage their computation. 

Microlensing in the Milky Way is a unique tool to break the dark-matter to baryonic matter degeneracy, however since a large fraction of the mass of the disk lies at these low latitudes measurements are essential in this challenging region. A further important tool to this goal will be microlensing surveys in the NIR. To help guide the design and assess the impact of future surveys we provide in \autoref{fig:fullmaps} maps of the optical depth across the entire Galactic bulge.

Recently \citet{Awiphan:16} also modelled the MOA-II data within the Besan\c on model, finding optical depths smaller than the observed data, particularly close to the Galactic plane. Some of this discrepancy was resolved by the revised MOA-II data, but the optical depths are remain lower than those observed. The dynamical models used in this work have higher bulge mass than those in \citet{Awiphan:16} which resolves the discrepancy. We note that while the bulge dynamical models do not include a nuclear bulge component inside $1\deg$  \citep{Launhardt:02}, since this is significant only inside $|b|<1\deg$ it does not effect the MOA-II measurements which lie outside this.

The optical depth is fundamentally a weighted mean of the density of lenses along the line-of-sight towards each source star (see \eg \autoref{eq:tau}). Because of this the microlensing data of the bulge constrains the stellar mass distribution towards centre of the Galaxy. The weighting of this density measurement is most sensitive to events halfway between the source and lens, and because of our position in the Galaxy is weighted to be sensitive to the stellar density at $R \sim 4\kpc$. Since this is likely to be near the position of the peak in the stellar rotation it makes the microlensing optical depth an attractive method to measure the disk maximality. A disadvantage is that since there is effectively only one line-of-sight (that towards the bulge) assumptions on the form of the density are needed to convert the optical depth to a stellar density constraint. In this work we assume the disk is axisymmetric, however this assumption is likely to be violated to some extent, and certainly inside the radius of the Galactic bar of $(5.0\pm0.2)\kpc$ \citep{Wegg:15}. Our fiducial model has a disk mass between 2.2\kpc and 5\kpc of $2.1\times10^{10}\msun$ while the mass of the bar outside the bulge derived by \citet{Wegg:15} is $1\times10^{10}\msun$. As a result we expect a significant fraction of the mass in this range may be part of the bar and therefore non-axisymmetric. Fortunately our line-of-sight to the bulge lies at an angle $(27\pm 2)\deg$ to the bar \citep{Wegg:13}, and as a result we are neither looking along or perpendicular to the bar. In addition the microlensing optical depth could be sensitive to the presence of a ring or trailing structures at the end of the bar, since this would lie almost halfway to the bulge, where the microlensing is most sensitive to the density of lenses.

We note that microlensing models of the bulge, such as that presented here, are important not just for the constraints they provide on the galaxy. Planet detection through microlensing has become as important tool to probe distant and low mass planets, and this is an important component of the upcoming EUCLID and WFIRST missions \citep{Beaulieu:11}. The expected yields and targeting requires microlensing models such as those described in this work.

\section{Conclusions}
\label{sec:conc}

Our fiducial model does an excellent job of predicting the optical depth, rates, and timescale distributions of both the MOA-II and EROS-II microlensing data. This model, constructed by adding an axisymmetric disk to the N-body bulges of \citet{Portail:15}, has a disk scale length of 2.6\kpc and a bulge dark matter fraction of 12\%. 

Our dynamical models also match the timescale distribution of microlensing events in the OGLE-III survey  very well for a \citet{Kroupa:01} or particularly a \citet{Calamida:15} log-normal IMF. Preliminary investigations suggest that a low number of brown dwarfs is required to in order to not overestimate the number of short duration events in the OGLE-III survey.

Microlensing in the Milky Way is a unique tool for breaking the degeneracy between dark matter and stellar mass-to-light. It is sensitive to the stellar density in the inner Milky Way where the dark matter contribution is most uncertain, and where the interplay between baryonic and dark matter are expected to be strongest. 

By varying the bulge stellar to dark matter fraction and the disk scale length and height we find a range of models consistent with the revised MOA-II data. In particular there is a degeneracy between the amount of stellar matter in the bulge and in the foreground disk. However the resultant rotation curves allowed by our models which match the microlensing data require a high baryonic fraction and permit only a limited amount of dark matter in the inner galaxy. The disk maximality, defined as the baryonic contribution to the rotation curve at its peak, is \onesigbounds\ at $1\sigma$, and $>\twosigmax$ at $2\sigma$. Maximal disk fits in external galaxies find $(0.85\pm0.10)$ \citep{Sackett:97} and so this places the Milky Way near the boundary between maximal and sub-maximal disks.

These high baryonic fractions in the inner Galaxy, and high levels of disk maximality are consistent with the NFW profiles predicted by pure $\Lambda$CDM simulations of Milky Way like halos. The $1\sigma$ bounds from the revised MOA-II data are inconsistent with strongly adiabatically contracted haloes. However more recent results suggest less strongly adiabatically haloes. Together with lower concentrations in line with the expected halo-to-halo concentration scatter produces haloes consistent with the data without the need for strong feedback altering the inner profiles of Milky Way sized haloes. 

Our results are driven by the optical depths in the revised MOA-II sample. The results in this work could be confirmed and improved though the estimation of optical depths from the larger OGLE-III survey, particularly close to the Galactic plane where much of the mass of the inner disk lies.



\bsp
\label{lastpage}
\end{document}